\documentclass[review,12pt,authoryear]{elsarticle}

\usepackage{amssymb}
\usepackage{amsthm}
\usepackage{amsmath}
\usepackage{mathrsfs}
\usepackage{graphicx}
\usepackage{epstopdf}
\usepackage{float}
\usepackage{caption}
\usepackage{subcaption}
\usepackage{bm}
\usepackage{bbm}
\usepackage{mathrsfs}
\usepackage{cleveref}
\usepackage{soul}
\usepackage{tikz} 
\usetikzlibrary{shapes.geometric, arrows} 
\biboptions{sort&compress}

\tikzstyle{startstop} = [rectangle, rounded corners, minimum width=1cm, minimum height=1cm,text centered, draw=black, fill=red!5]
\tikzstyle{io} = [trapezium, trapezium left angle=70, trapezium right angle=110, minimum width=1cm, minimum height=1cm, text centered, draw=black, fill=blue!5]
\tikzstyle{process} = [rectangle, minimum width=1cm, minimum height=1cm, text centered, draw=black, fill=orange!5]
\tikzstyle{decision} = [diamond, minimum width=1cm, minimum height=0.5cm, text centered, draw=black, fill=green!5]
\tikzstyle{arrow} = [thick,->,>=stealth]

\journal{Applications in Engineering Science}

\makeatletter
\def\@author#1{\g@addto@macro\elsauthors{\normalsize%
    \def\baselinestretch{1}%
    \upshape\authorsep#1\unskip\textsuperscript{%
      \ifx\@fnmark\@empty\else\unskip\sep\@fnmark\let\sep=,\fi
      \ifx\@corref\@empty\else\unskip\sep\@corref\let\sep=,\fi
      }%
    \def\authorsep{\unskip,\space}%
    \global\let\@fnmark\@empty
    \global\let\@corref\@empty  
    \global\let\sep\@empty}%
    \@eadauthor={#1}
}
\makeatother

\makeatletter
\def\thickhline{%
  \noalign{\ifnum0=`}\fi\hrule \@height \thickarrayrulewidth \futurelet
   \reserved@a\@xthickhline}
\def\@xthickhline{\ifx\reserved@a\thickhline
               \vskip\doublerulesep
               \vskip-\thickarrayrulewidth

             \fi
      \ifnum0=`{\fi}}
\makeatother

\newlength{\thickarrayrulewidth}
\begin{document}

\begin{frontmatter}



\title{A simple and robust Abaqus implementation of the phase field fracture method}


\author{Yousef Navidtehrani\fnref{Uniovi}}

\author{Covadonga Beteg\'{o}n\fnref{Uniovi}}

\author{Emilio Mart\'{\i}nez-Pa\~neda\corref{cor1}\fnref{IC}}
\ead{e.martinez-paneda@imperial.ac.uk}

\address[Uniovi]{Department of Construction and Manufacturing Engineering, University of Oviedo, Gij\'{o}n 33203, Spain}

\address[IC]{Department of Civil and Environmental Engineering, Imperial College London, London SW7 2AZ, UK}

\cortext[cor1]{Corresponding author.}

\begin{abstract}
The phase field fracture method is attracting significant interest. Phase field approaches have enabled predicting - on arbitrary geometries and dimensions - complex fracture phenomena such as crack branching, coalescence, deflection and nucleation. In this work, we present a simple and robust implementation of the phase field fracture method in the commercial finite element package Abaqus. The implementation exploits the analogy between the phase field evolution law and the heat transfer equation, enabling the use of Abaqus' in-built features and circumventing the need for defining user elements. The framework is general, and is shown to accommodate different solution schemes (staggered and monolithic), as well as various constitutive choices for preventing damage under compression. The robustness and applicability of the numerical framework presented is demonstrated by addressing several 2D and 3D boundary value problems of particular interest. Focus is on the solution of paradigmatic case studies that are known to be particularly demanding from a convergence perspective. The results reveal that our phase field fracture implementation can be readily combined with other advanced computational features, such as contact, and deliver robust and precise solutions. The code developed can be downloaded from www.empaneda.com/codes.
\end{abstract}


\end{frontmatter}


\section{Introduction}
\label{Introduction}

Modelling the morphology of an evolving interface is considered to be a longstanding mathematical and computational challenge. Tracking interface boundaries explicitly is hindered by the need of defining moving interfacial boundary conditions and manually adjusting the interface topology with arbitrary criteria when merging or division occurs \citep{Biner2017}. Phase field formulations have proven to offer a pathway for overcoming these challenges. In the phase field modelling paradigm, the interface is smeared over a \emph{diffuse} region using an auxiliary field variable $\phi$, which takes a distinct value for each of the two phases (e.g., 0 and 1) and exhibits a smooth change between these values near the interface. The temporal evolution of the phase field variable $\phi$ is described by a partial differential equation (PDE) and thus the method enables the simulation of complex interface evolution phenomena by integrating a set of PDEs for the whole system, avoiding the explicit treatment of interface conditions.\\

The phase field paradigm has quickly gained significant traction in the condensed matter and materials science communities, becoming the \textit{de facto} tool for modelling microstructural evolution \citep{Provatas2011}. The change in shape and size of microstructural features such as grains can be predicted by defining the evolution of the phase field in terms of other fields (temperature, concentration, strain, etc.) through a thermodynamic free energy. This success has been extended to other interfacial problems, such as corrosion, where the phase field smoothens the metal-electrolyte interface \citep{JMPS2021}, or fracture mechanics, where the phase field is used to implicitly track the evolution of the crack-solid boundary \citep{Bourdin2000}. The coupling of the phase field paradigm with the variational approach to fracture presented by \citet{Bourdin2008} has opened new horizons in the modelling of cracking phenomena, from predicting complex crack trajectories to simulating inertia-driven crack branching. Moreover, this can be achieved on the original finite element mesh, without \textit{ad hoc} crack propagation criteria, and for arbitrary geometries and dimensions. Not surprisingly, the popularity of phase field methods for fracture has rocketed in recent years; applications include the prediction of fracture (and fatigue) in fibre-reinforced composites \citep{Quintanas-Corominas2019,CST2021}, hydrogen-embrittled alloys (\citealp{CMAME2018}; \citealp{JMPS2020}), batteries \citep{Miehe2015,Klinsmann2016a}, rock-like materials (\citealp{Zhou2019b}; \citealp{Schuler2020}), solar-grade silicon \citep{Paggi2018}, functionally graded materials \citep{CPB2019,Kumar2021}, hyperelastic solids \citep{Loew2019,Mandal2020a}, piezo-electric materials \citep{Abdollahi2012} and shape memory alloys \citep{CMAME2021} - see \citep{Wu2020} for a comprehensive review.\\ 

The success of phase field fracture methods has also triggered a notable interest for the development of robust solution algorithms to solve the coupled deformation-fracture problem (\citealp{Miehe2010}; \citealp{Gerasimov2016}; \citealp{Wu2020a}; \citealp{TAFM2020}). The total potential energy functional, including the contributions from the bulk and fracture energies, is minimised with respect to the two primary kinematic variables: the displacement field $\bm{u}$ and the phase field $\phi$. Thus, the phase field $\phi$, a damage-like variable, is solved for at the finite element nodes, as an additional degree of freedom. This requires performing the numerical implementation at the element level, as opposed to local damage models, which are implemented at the integration point level. In the context of commercial finite element packages, solving for the phase field as a degree-of-freedom requires the development of user element subroutines. The commercial finite element package Abaqus has received particular attention in the phase field fracture community, and a vast literature has emerged on the implementation of the phase field fracture method on this popular software suite (\citealp{Liu2016a}; \citealp{Molnar2017}; \citealp{Fang2019}; \citealp{Molnar2020}; \citealp{Wu2020c}). These implementations require programming an \textit{ad hoc} finite element, effectively using Abaqus as a solver and not being able to exploit most of its in-built features. In this work, we circumvent this issue by exploiting the analogy between the heat conduction equation and the phase field evolution law. This approach enables using the vast majority of Abaqus' in-built features, including the coupled temperature-displacement elements from its finite element library, which avoids coding user-defined elements and the associated complications in meshing and visualisation (e.g., Abaqus2Matlab is frequently used to pre-process input files, \citealp{AES2017}). Moreover, the phase field implementation presented can accommodate both staggered and monolithic solution schemes, ensuring convergence in all cases. We demonstrate the potential and robustness of the implementation presented by addressing several paradigmatic 2D and 3D boundary value problems. The framework provided is general and can be easily implemented in other finite element packages. \\

The remainder of this manuscript is organised as follows. In Section \ref{Sec:PhaseField} we describe the theory underlying the phase field fracture method. The analogy with the heat transfer problem and the implementation details are given in Section \ref{Sec:Implementation}. Representative results are shown in Section \ref{sec:Results}. First, unstable fracture is addressed with the paradigmatic benchmark of a cracked square plate under uniaxial tension. Secondly, convergence under stable crack propagation conditions is investigated using a cracked square plate subjected to shear. The performance of monolithic and staggered schemes is compared. Thirdly, the screw tension tests presented by \citet{Wick2015} are examined. Finally, we simulate the so-called Brazilian laboratory test, which is widely used for measuring the tensile strength of rock-like materials. A comprehensive 3D analysis is conducted, including the modelling of the contact between the jaws and the specimen. The manuscript ends with concluding remarks in Section \ref{Sec:ConcludingRemarks}. 

\section{Phase field fracture model}
\label{Sec:PhaseField}

The phase field fracture method builds upon Griffith's thermodynamics framework \citep{Griffith1920}. In agreement with the first law of thermodynamics, a crack can form (or grow) only if this process causes the total energy of the system to decrease or remain constant. Accordingly, a critical condition for fracture can be defined upon the assumption of equilibrium conditions - no net change in total energy. Consider an elastic solid containing a crack. In the absence of external forces, the variation of the total energy $\mathcal{E}$ due to an incremental increase in the crack area d$A$ is given by
\begin{equation}\label{eq:Egriffith0}
    \frac{\text{d}\mathcal{E}}{\text{d}A} = \frac{\text{d} \psi \left( \bm{\varepsilon} \left( \bm{u} \right) \right) }{\text{d} A} + \frac{\text{d}W_c}{\text{d}A} = 0
\end{equation}

\noindent where $W_c$ is the work required to create new surfaces and $\psi$ is the strain energy density, which is a function of the displacement field $\bm{u}$ and the strain field $\bm{\varepsilon}=\left( \nabla \bm{u}^T + \nabla \bm{u} \right)/2$. The last term in Eq. (\ref{eq:Egriffith0}) is the so-called critical energy release rate $G_c=\text{d}W_c/\text{d}A$, a material property that characterises the fracture resistance. Thus, Griffth's premise is a local minimality principle for the sum of the elastic and fracture energies. For an arbitrary body $\Omega \subset {\rm I\!R}^n$ $(n \in[1,2,3])$ with internal discontinuity boundary $\Gamma$, this minimality principle can be expressed in a variational form as \citep{Bourdin2008},
\begin{equation}\label{eq:Egriffith}
    \mathcal{E} \left( \bm{u} \right) = \int_\Omega  \psi \left( \bm{\varepsilon} \left( \bm{u} \right) \right) \, \text{d}V +  \int_\Gamma  G_c \, \text{d}S \, ,
\end{equation}

Thus, within this framework, crack growth along any trajectory can be predicted without arbitrary criteria, driven by global minimality and the transformation of stored energy into fracture energy. However, minimisation of the variational Griffith energy functional (\ref{eq:Egriffith}) is hindered by the complexities associated with tracking the propagating fracture surface $\Gamma$. The problem can be made computationally tractable by employing an auxiliary phase field $\phi$ that enables tracking the crack interface. The phase field $\phi$ can be interpreted as a damage-like variable that goes from 0 in intact regions to 1 inside of the crack. Accordingly, following continuum damage mechanics arguments, a degradation function $g (\phi)=(1-\phi)^2$ can be defined to reduce the material stiffness with evolving damage. Hence, the regularised energy functional is given by,
\begin{equation}\label{eq:Egriffith1}
    \mathcal{E}_\ell \left( \bm{u}, \phi \right) = \int_\Omega \left( 1 - \phi \right)^2 \psi_0 \left( \bm{\varepsilon} \left( \bm{u} \right) \right) \, \text{d}V + \int_\Omega G_c  \gamma_\ell \left( \phi \right) \, \text{d}V \, ,
\end{equation}

\noindent where $\ell$ is a length scale parameter that governs the size of the fracture process zone and $\gamma_\ell$ is the crack density function. A common choice for $\gamma_\ell$ reads,
\begin{equation}
    \gamma_\ell \left( \phi \right) = \frac{\phi^2}{2 \ell} + \frac{\ell}{2} |\nabla \phi|^2 \, .
\end{equation}

As rigorously proven using Gamma-convergence, the $(\bm{u}, \phi)$ sequence that constitutes a global minimum for the regularised functional $\mathcal{E}_\ell$ converges to that of $\mathcal{E}$ for a fixed $\ell \to 0^+$. Thus, $\ell$ can be interpreted as a regularising parameter in its vanishing limit. However, for $\ell>0^+$ a finite material strength is introduced and thus $\ell$ becomes a material property governing the strength \citep{Tanne2018}; e.g., for plane stress:
\begin{equation}
  \sigma_f  \propto \sqrt{\frac{G_c E}{\ell}} = \frac{K_{Ic}}{\sqrt{\ell}}
\end{equation}
\noindent where $K_{Ic}$ is the material fracture toughness. It has been shown that the consideration of a finite $\ell>0^+$ enables to accurately predict crack nucleation, capturing its transition from strength-driven to fracture-driven \citep{Tanne2018}, and in agreement with the predictions from the coupled criterion in finite fracture mechanics \citep{Molnar2020a}.\\

We will restrict our analysis to the behaviour of linear elastic materials, such that the strain energy density of the intact material is given by,
\begin{equation}
    \psi_0 = \frac{1}{2} \bm{\varepsilon} : \bm{C}_0 : \bm{\varepsilon},
\end{equation}

\noindent where $\bm{C}_0$ is the (undamaged) linear elastic stiffness tensor. Accordingly, the Cauchy stress tensor is defined as 
\begin{equation}\label{eq:Cauchy}
    \bm{\sigma} = \left( 1 - \phi \right)^2 \bm{\sigma}_0 = \left( 1 - \phi \right)^2 \frac{\partial \psi_0 \left( \bm{\varepsilon} \right) }{\partial \bm{\varepsilon}}
\end{equation}

\noindent where the undamaged Cauchy stress is given by $\bm{\sigma}_0 =\bm{C}_0 : \bm{\varepsilon}$.\\

Considering the constitutive choices just described and taking the first variation of the $\mathcal{E}_\ell$ with respect to the primal kinematic variables $\bm{u}$ and $\phi$ renders,
\begin{align}\label{eq:weakform}
    \int_\Omega  \bigg[ \left( 1 - \phi \right)^2 \bm{\sigma}_0 : \text{sym} \nabla \delta \bm{u} & - 2 \left( 1 - \phi \right)  \, \psi_0 \left( \bm{\varepsilon} \left( \bm{u} \right) \right) \delta \phi \nonumber \\ 
    & + G_c \left( \frac{\phi}{\ell} \delta \phi + \ell \nabla \phi \cdot \nabla \delta \phi \right) \bigg] \, \text{d} V =0
\end{align}

The local force balances can be readily derived by applying Gauss' divergence theorem and noting that (\ref{eq:weakform}) must hold for any kinematically admissible variations of the virtual quantities. Thus, the coupled field equations read, 
\begin{align}\label{eqn:strongForm}
\nabla \cdot  \left[ (1-\phi)^2 \boldsymbol{\sigma}_0 \right]  &= \boldsymbol{0}   \hspace{3mm} \rm{in}  \hspace{3mm} \Omega \nonumber \\ 
G_{c}  \left( \dfrac{\phi}{\ell}  - \ell \Delta \phi \right) - 2(1-\phi) \, \psi_0 \left( \bm{\varepsilon} \left( \bm{u} \right) \right) &= 0 \hspace{3mm} \rm{in} \hspace{3mm} \Omega  
\end{align}

\noindent The discretised forms of the field equations can be solved using a monolithic scheme, where $\bm{u}$ and $\phi$ are solved simultaneously, or by means of a so-called staggered scheme, where an alternate minimisation strategy is used.

\section{Finite element implementation}
\label{Sec:Implementation}

We shall describe the numerical framework proposed. First, we introduce a history field to ensure damage irreversibility. Secondly, the analogy with heat transfer is presented. Thirdly, the particularities of the Abaqus implementation are described. Finally, we show how our implementation can accommodate different solution schemes, and discuss the advantages and limitations of the options available. For the sake of brevity, we limit our description to the constitutive and implementation choices inherent to the code provided, and describe in \ref{App:FEM} other potential extensions, which are considered in the numerical examples.

\subsection{Damage irreversibility}
\label{Sec:DamageIrreversibility}

A history variable field $H$ is introduced to prevent crack healing, ensuring that the following condition is always met
\begin{equation} \label{growth}
    \phi_{t+\Delta t} \geq \phi_{t} \, ,
    \centering
\end{equation}

\noindent where $\phi_{t+\Delta t}$ is the phase field variable in the current time increment while $\phi_{t}$ denotes the value of the phase field on the previous increment. For both loading and unloading scenarios, the history field must satisfy the Kuhn-Tucker conditions
\begin{equation}
    \psi_{0} - H \leq 0 \text{,} \hspace{7mm} \dot{H} \geq 0 \text{,} \hspace{7mm} \dot{H}(\psi_{0}-H)=0 \, .
    \centering
\end{equation}
\noindent Accordingly, the history field for a current time $t$ can be written as: 
\begin{equation}\label{eq:History}
     H = \max_{\tau \in[0,t]}\psi_0( \tau). 
\end{equation}

\subsection{Heat Transfer Analogy}

For a solid with thermal conductivity $k$, specific heat $c_p$ and density $\rho$, the field equation for heat transfer in the presence of a heat source $r$ reads:
\begin{equation}\label{eq:HeatTransfer1}
    k \nabla^2 T - \rho c_p \frac{\partial T}{\partial t} = r \, ,
\end{equation}

\noindent where $T$ is the temperature field. Under steady-state conditions the rate term vanishes and Eq. (\ref{eq:HeatTransfer1}) is reduced to,
\begin{equation}\label{eq:HeatTransfer2}
    k \nabla^2 T = r
\end{equation}

The analogy of this elliptic partial differential equation (PDE) with the phase field evolution law is evident, with the temperature field acting as the phase field $T \equiv \phi$. Making use of the history field described above, one can reformulate the phase field local force balance, Eq. (\ref{eqn:strongForm})b, as
\begin{equation}\label{eq:HeatTransferPhase}
    \nabla^2 \phi = \frac{\phi}{\ell^2} - \frac{2 \left( 1 - \phi \right)}{G_c \ell} H \, .
\end{equation}

\noindent And thus (\ref{eq:HeatTransfer2}) and (\ref{eq:HeatTransferPhase}) are equivalent upon assigning the value of unity to the thermal conductivity ($k=1$) and defining the following heat flux due to internal heat generation,
\begin{equation}\label{eq:r}
    r = \frac{\phi}{\ell^2} - \frac{2 \left( 1 - \phi \right)}{G_c \ell} H \, .
\end{equation}

Finally, for the computation of the Jacobian matrix, one should also define the rate of change of heat flux ($r$) with temperature ($T\equiv \phi$),
\begin{equation}\label{eq:r_phi}
    \frac{\partial r}{\partial \phi} = \frac{1}{\ell^2} + \frac{2 H}{G_c \ell}
\end{equation}

We have restricted ourselves to the steady-state scenario, treating the phase field evolution law as rate-independent. This is, by far, the most common formulation for phase field fracture. However, one can also introduce a viscous regularisation term in the phase field equation by exploiting instead the transient problem - Eq. (\ref{eq:HeatTransfer1}). In such scenario, the quantity $\rho c_p$ is analogous to a viscosity parameter \citep{Miehe2010a}. The heat capacity terms help stabilising the solution and thus one might wish to address a rate-independent (steady-state) problem by conducting instead a transient analysis over a long time. However, as demonstrated in the numerical examples below, we do not see the need to consider viscous regularisation to achieve convergence. 

\subsection{Abaqus particularities}

\begin{figure}[H]
\centering
\noindent\makebox[\textwidth]{%
\includegraphics[scale=0.2]{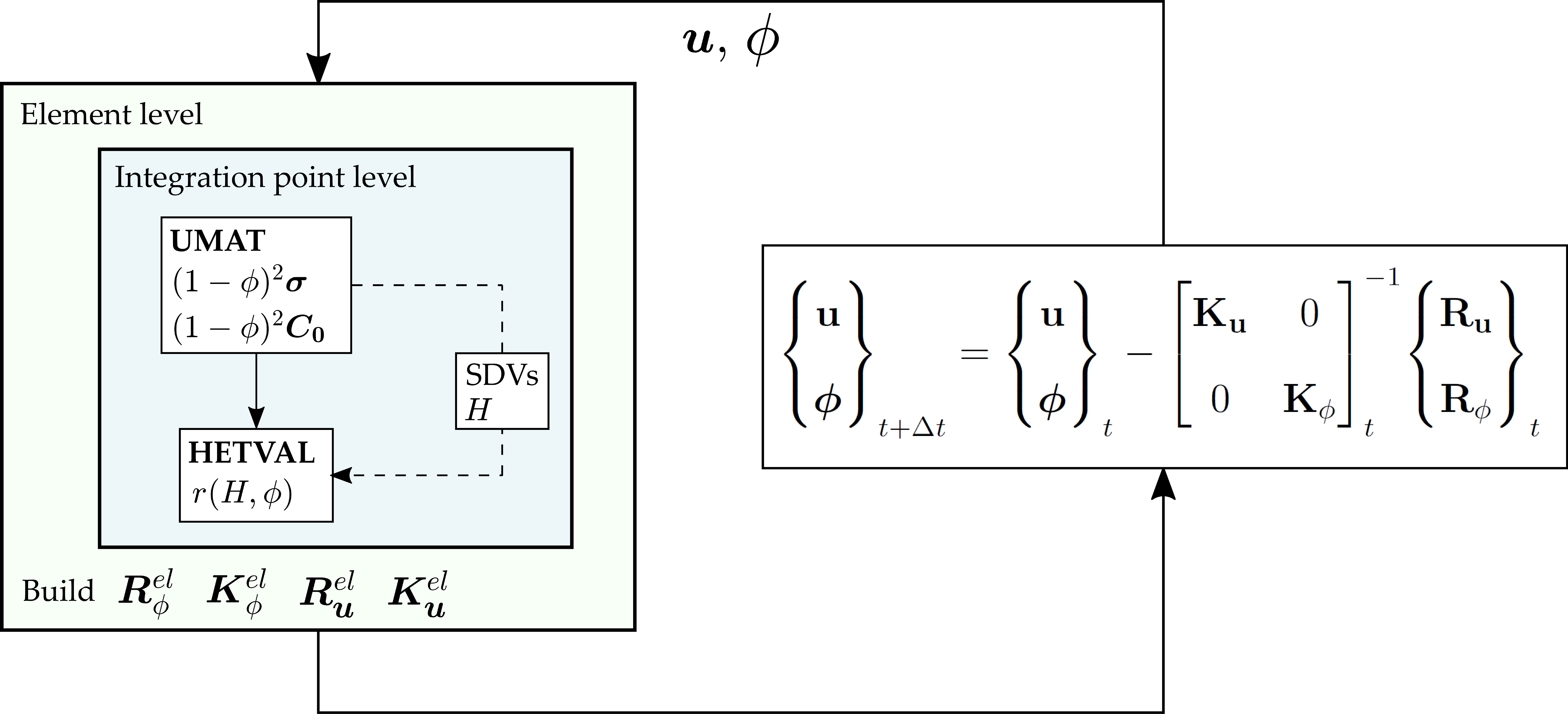}}
\caption{User subroutine flowchart for the implementation of a coupled deformation - phase field fracture model exploiting the analogy with heat transfer.}
\label{Fig:UMATHETVAL}
\end{figure}

The heat transfer analogy described can be readily implemented in Abaqus by making use of user material (UMAT) and heat flux (HETVAL) subroutines. The process is outlined in Fig. \ref{Fig:UMATHETVAL}. Taking advantage of the heat transfer analogy enables carrying out the implementation at the integration point level, using in-built displacement-temperature elements such as the Abaqus CPE4T type for the case of 4-node bilinear quadrilateral elements. For a given element, Abaqus provides to the integration point-level subroutines the values of strain and phase field (temperature), as interpolated from the nodal solutions. Within each integration point loop, the user material subroutine (UMAT) is called first. Inside of the UMAT, the material Jacobian $\bm{C}_0$ and the Cauchy stress $\bm{\sigma}$ can be readily computed from the strain tensor. The current value of the phase field (temperature) is then used to account for the damage degradation of these two quantities. The strain energy density can be stored in so-called solution dependent state variables (SDVs), enabling to enforce the irreversibility condition (Section \ref{Sec:DamageIrreversibility}). The updated value of the SDVs is transferred to the heat flux (HETVAL) subroutine; this is used to transfer the current value of the history field $H$, without the need for external Fortran modules. In the HETVAL subroutine we define the internal heat flux $r$, Eq. (\ref{eq:r}), and its derivative with respect to the temperature (phase field) $\partial r/ \partial \phi$, Eq. (\ref{eq:r_phi}). The process is repeated for every integration point, enabling Abaqus to externally build the element stiffness matrices and residuals and assembling the global system of equations, see Fig. \ref{Fig:UMATHETVAL}. It is worth emphasising that the coupling terms in the stiffness matrix are not defined: $\bm{K}_{\bm{u}\phi}=\bm{K}_{\phi \bm{u}}=\bm{0}$, making the stiffness matrix symmetric. By default, Abaqus assumes a non-symmetric system for coupled displacement-temperature analyses but this can be modified by defining a separated solution technique. It should be noted that parallel calculations using versions of Abaqus older than 2016 only execute the solver in parallel (if the separated solution technique is used).

To avoid editing the user subroutine, mechanical and fracture properties are defined in the input file only, as user material properties, and are then transferred between subroutines using solution dependent variables. Consistent with the heat transfer analogy outlined above, one must activate the heat generation option and define as material properties the thermal conductivity $k$, the density $\rho$ and the specific heat $c_p$, all of them with a value of unity. Also, one should  assign an initial temperature distribution of $T (t=0)=0 \,  \forall \, \bm{x}$. No additional pre-processing or post-processing steps are needed, all actions can be conducted within the Abaqus/CAE graphical user interface and the phase field solution can be visualised by plotting the nodal solution temperature (NT11). 

\subsection{Solution schemes}

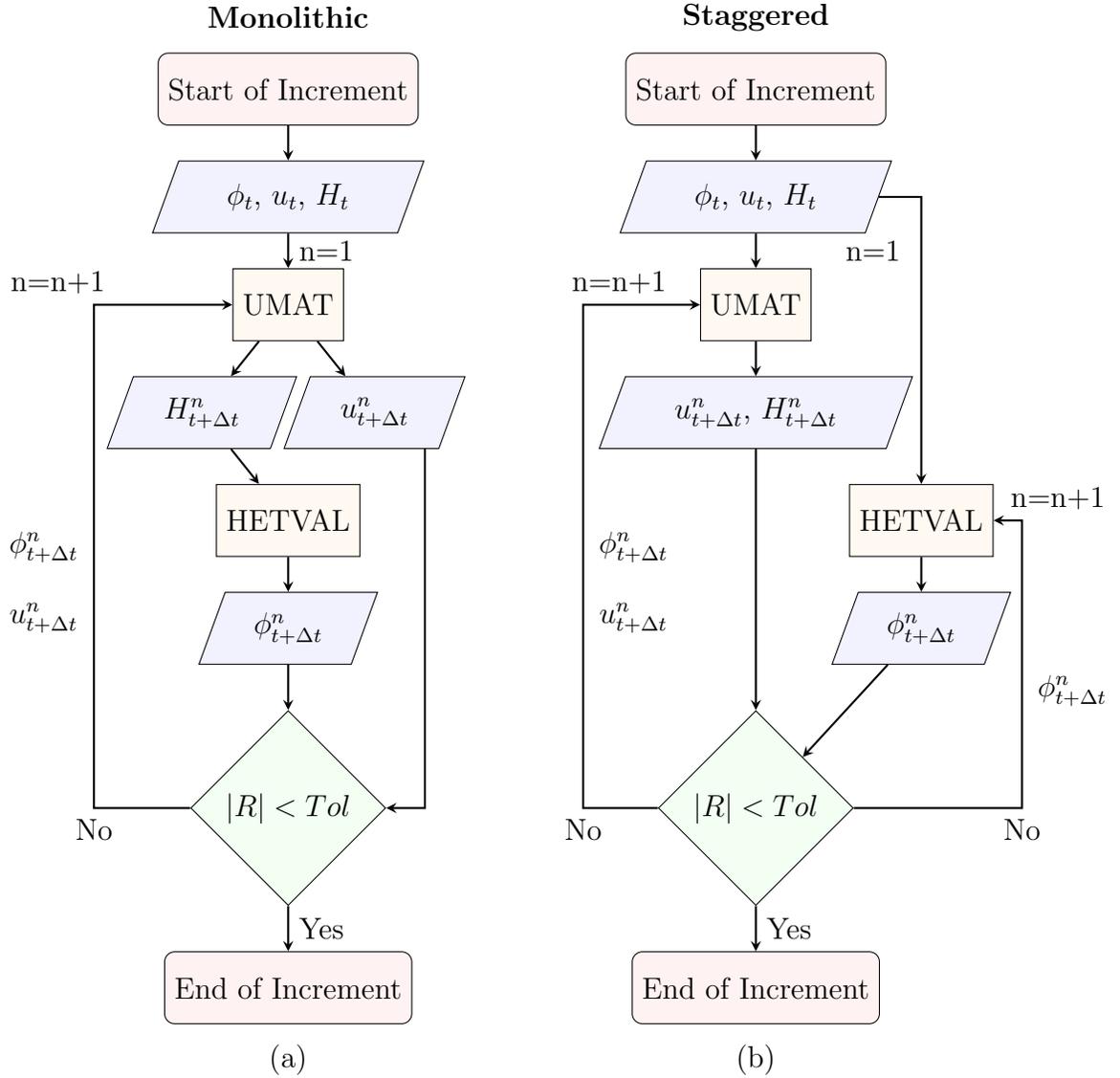
\begin{figure}[H]
    \begin{tikzpicture}[node distance=1.5cm]
    \node (start) [startstop] {Start of Increment};
    \node (ti1) [yshift=1cm] {\textbf{Monolithic}};
    \node (in1) [io, below of=start] {$\phi_t$, $u_t$, $H_t$};
    \node (pro1) [process, below of=in1] {UMAT};
    \node (in2) [io, below of=pro1, xshift=1.2cm] {$u_{t+\Delta t}^{n}$};
    \node (in3) [io, below of=pro1, xshift=-1.2cm] {$H_{t+\Delta t}^{n}$};
    \node (pro2) [process, below of=in3, xshift=1.2cm] {HETVAL};
    \node (out1) [io, below of=pro2] {$\phi_{t+\Delta t}^{n}$};
    \node (dec1) [decision, below of=out1, yshift=-1cm] {$|R| < Tol$};
    \node (end) [startstop, below of=dec1, yshift=-1cm] {End of Increment};
    \node (cap) [below of=end, yshift=.5cm] {(a)};
        
    \draw [arrow] (start) -- (in1);
    \draw [arrow] (in1) -- node[anchor=west] {n=1} (pro1);
    \draw [arrow] (pro1) -- (in2);
    \draw [arrow] (pro1) -- (in3);
    \draw [arrow] (in3) -- (pro2);
    \draw [arrow] (pro2) -- (out1);
    \draw [arrow] (out1) -- (dec1);
    \draw [arrow] (in2) +(.7,-.5) |- (dec1);
    \draw [arrow] (dec1) -- +(-2.7,0) node[anchor=north] {No}
    node[anchor=north, xshift=-.7cm, yshift=3cm] {$u_{t+\Delta t}^{n}$}
    node[anchor=north, xshift=-.7cm, yshift=4cm] {$\phi_{t+\Delta t}^{n}$}|- node[anchor=south, xshift=-0.5cm] {n=n+1} (pro1);
    \draw [arrow] (dec1) -- node[anchor=west] {Yes} (end);

    \node (start-2) [startstop, right of=start, xshift=5cm] {Start of Increment};
    \node (ti2) [right of=ti1, xshift=5cm] {\textbf{Staggered}};
    \node (in1-2) [io, below of=start-2] {$\phi_t$, $u_t$, $H_t$};
    \node (pro1-2) [process, below of=in1-2] {UMAT};
    \node (in2-2) [io, below of=pro1-2] {$u_{t+\Delta t}^{n}$, $H_{t+\Delta t}^{n}$};
    \node (pro2-2) [process, below of=in2-2, xshift=2.3cm] {HETVAL};
    \node (out1-2) [io, below of=pro2-2] {$\phi_{t+\Delta t}^{n}$};
    \node (dec1-2) [decision, below of=in2-2, yshift=-4cm] {$|R| < Tol$};
    \node (end-2) [startstop, below of=dec1-2, yshift=-1cm] {End of Increment};
    \node (cap-2) [below of=end-2, yshift=.5cm] {(b)};

    \draw [arrow] (start-2) -- (in1-2);
    \draw [arrow] (in1-2) -- node[anchor=west, xshift=1.1cm] {n=1} (pro1-2);
    \draw [arrow] (pro1-2) -- (in2-2);
    \draw [arrow] (in1-2) -| (pro2-2);
    \draw [arrow] (pro2-2) -- (out1-2);
    \draw [arrow] (out1-2) --  (dec1-2);
    \draw [arrow] (in2-2) --  (dec1-2);
    \draw [arrow] (dec1-2) -- node[anchor=west] {Yes} (end-2);
    \draw [arrow] (dec1-2) -- +(-2.4,0) node[anchor=north] {No}
    node[anchor=north, xshift=.7cm, yshift=3cm] {$u_{t+\Delta t}^{n}$}
    node[anchor=north, xshift=.7cm, yshift=4cm] {$\phi_{t+\Delta t}^{n}$} |-
    node[anchor=south, xshift=0.5cm] {n=n+1} (pro1-2);
    \draw [arrow] (dec1-2) -- +(3.7,0) node[anchor=north] {No} 
    node[anchor=north, xshift=+.7cm, yshift=2cm] {$\phi_{t+\Delta t}^{n}$} |-
    node[anchor=south, xshift=0.5cm] {n=n+1} (pro2-2);
        \end{tikzpicture}
        \caption{Phase field fracture solution flowchart at each integration point for a specific increment: (a) monolithic, and (b) staggered schemes.}
        \label{Fig:Flowchart}
\end{figure}

The global system of equations, shown in Fig. \ref{Fig:UMATHETVAL}, can be solved in either a \emph{monolithic} or a \emph{staggered} manner. In a monolithic approach, the displacement sub-system $\bm{K}_{\bm{u}} \bm{u} = \bm{R}_{\bm{u}}$ and the phase field sub-system $\bm{K}_{\phi} \bm{\phi} = \bm{R}_{\phi}$ are solved simultaneously. On the other hand, a staggered solution scheme entails an alternative minimisation approach, by which the sub-systems are solved sequentially. Monolithic solution strategies are unconditionally stable and, therefore, more efficient (in principle). However, the total potential energy functional (\ref{eq:Egriffith1}) is non-convex with respect to $\bm{u}$ and $\phi$. As a consequence, the Jacobian matrix in Newton's method becomes indefinite, hindering convergence when solving for the displacement and the phase field at the same time. It has been recently shown that the use of quasi-Newton methods such as the Broyden-Fletcher-Goldfarb-Shanno (BFGS) algorithm enables the implementation of robust monolithic schemes that are very efficient and do not exhibit convergence issues 
\citep{Wu2020a,TAFM2020} - see also \citep{TAFM2021,Wu2021} for application examples. Unfortunately, the quasi-Newton solution scheme is not available in Abaqus for thermo-mechanical problems. Accordingly, we implement a conventional monolithic scheme, based on Newton's method, and a staggered scheme of the single-pass type. The flowchart associated with each of these solution schemes is presented in Fig. \ref{Fig:Flowchart}. In the staggered case, the residual and the stiffness matrix for the phase field sub-system are built considering the history field of the previous increment $H_{t}$; i.e., the history field is frozen during the iterative procedure, facilitating convergence in demanding problems at the cost of scarifying unconditional stability. A recursive iteration or multi-pass staggered scheme can be implemented by using a Fortran module to transfer the history field between the UMAT and the HETVAL. Thus, we provide a general framework that provides flexibility to enhance robustness or efficiency, as required for the problem at hand. This trade-off between efficiency and robustness, and the differences in performance between solution schemes, are addressed in the numerical examples below.\\

\section{Results}
\label{sec:Results}

We shall show the robustness and capabilities of the present implementation by simulating fracture in several paradigmatic boundary value problems. First, crack initiation and growth in a notched square plate is addressed under both uniaxial tension (Section \ref{Sec:PlateTension}) and shear (Section \ref{Sec:PlateShear}). Then, the failure of screws subjected to tension, with and without initial cracks, is simulated in Section \ref{Sec:Screw}. Finally, in Section \ref{Sec:Brazilian}, a 3D model of the Brazilian test is developed, including the contact between the jaws and the sample, to determine the nucleation and coalescence of cracks. 

\subsection{Notched square plate under tension}
\label{Sec:PlateTension}

\begin{figure}[H]
    \centering
    \begin{subfigure}[H]{0.45\textwidth}
        \includegraphics[width=\textwidth]{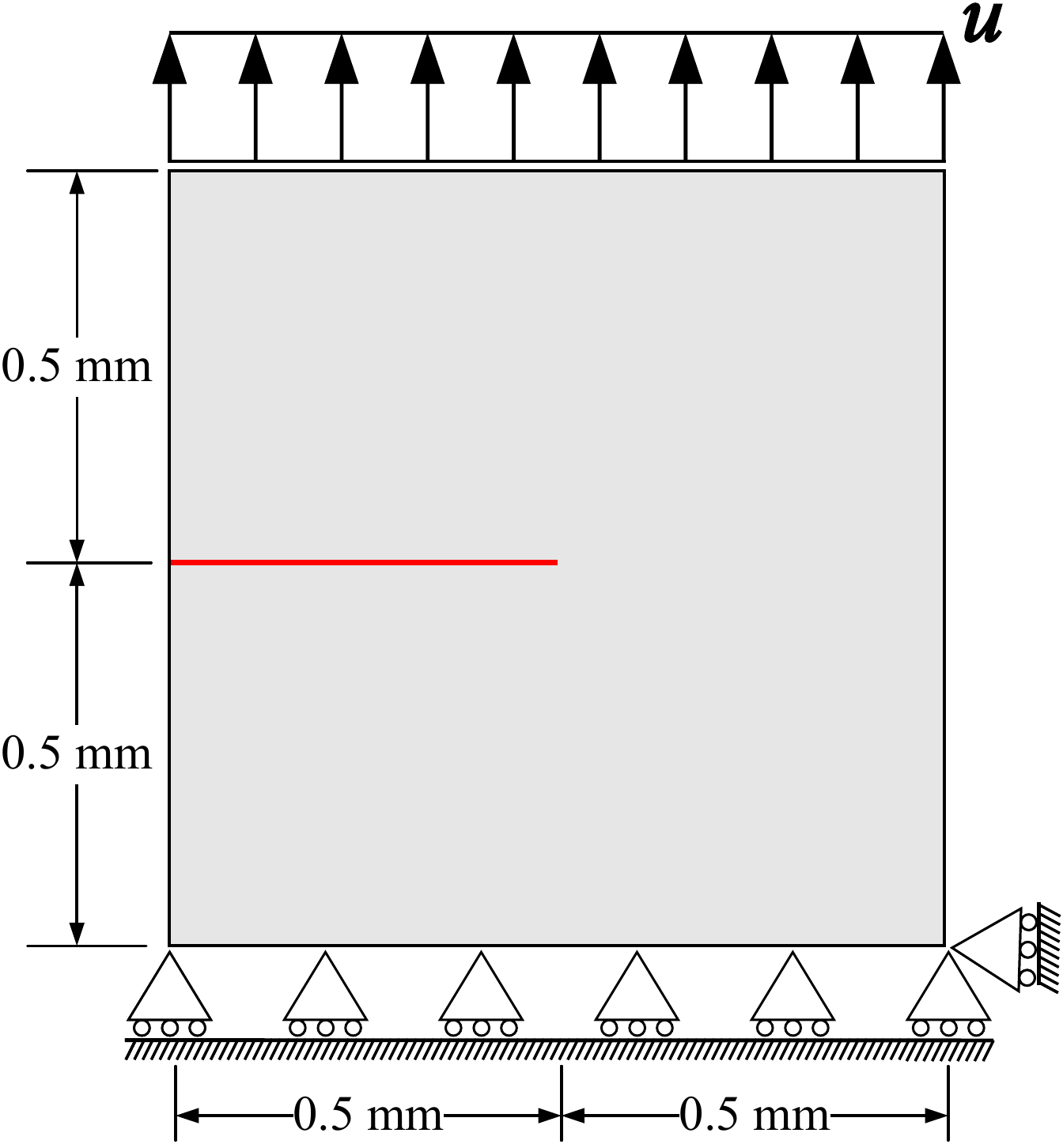}
        \caption{}
        \label{Fig:Sq-geo}
    \end{subfigure}
        \begin{subfigure}[H]{0.45\textwidth}
        \includegraphics[width=\textwidth]{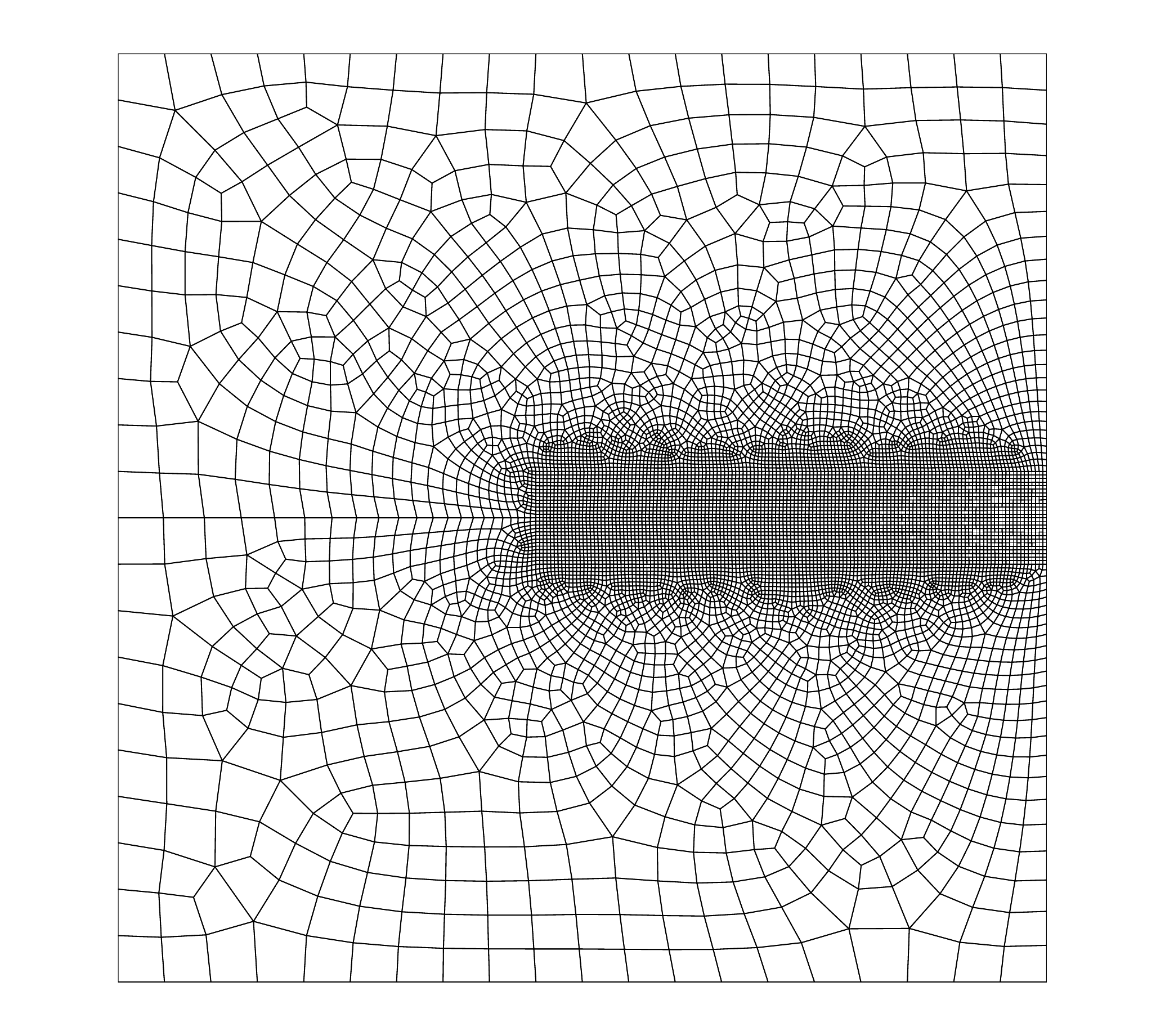}
        \caption{}
        \label{Fig:Sq-mesh}
    \end{subfigure}\hspace{0.1\textwidth}
    \begin{subfigure}[H]{0.45\textwidth}
        \includegraphics[width=\textwidth]{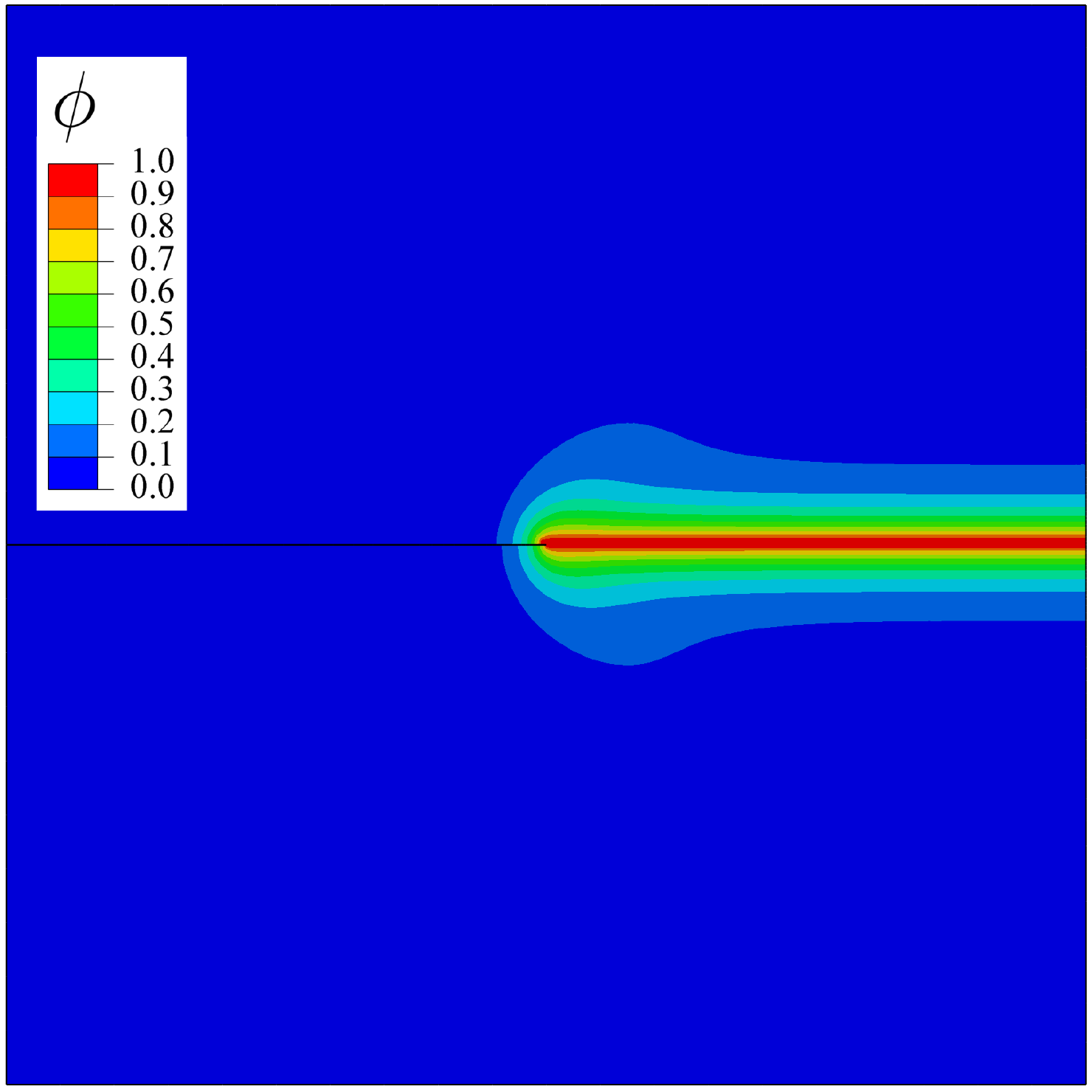}
        \caption{}
        \label{Fig:Sq-phi}
    \end{subfigure}
    \caption{Notched square plate under tension: (a) geometry, dimensions and boundary conditions, (b) finite element mesh, and (c) contour of the phase field $\phi$ after rupture.}
\end{figure}

First, we shall consider the case of unstable crack growth in a notched squared plate undergoing uniaxial tension. This is a paradigmatic benchmark in the phase field fracture community since the early work by \citet{Miehe2010}. The geometry and boundary conditions are shown in Fig. \ref{Fig:Sq-geo}. The sample is subjected to mode I fracture conditions, with a vertical displacement being prescribed in the remote boundary. The mechanical behaviour is characterised by a Young's modulus $E=210$ GPa and a Poisson's ratio $\nu=0.3$, while the fracture properties read $\ell=0.024$ mm and $G_c = 2.7$ N/mm \citep{TAFM2020}. We discretise the model using linear quadrilateral elements for coupled displacement-thermal analyses, CPE4T in Abaqus terminology. A total of 8,532 elements are used. As shown in Fig. \ref{Fig:Sq-mesh}, the mesh is refined along the expected crack path, such that the characteristic element size is at least five times smaller than the phase field length scale $\ell$. For this case study, the monolithic implementation is used and no strain energy decomposition is assumed. The predicted crack path is showcased in Fig. \ref{Fig:Sq-phi} by plotting the contours of the phase field variable $\phi$.    

\begin{figure}[H]
    \centering
    \includegraphics[width=1.0\textwidth]{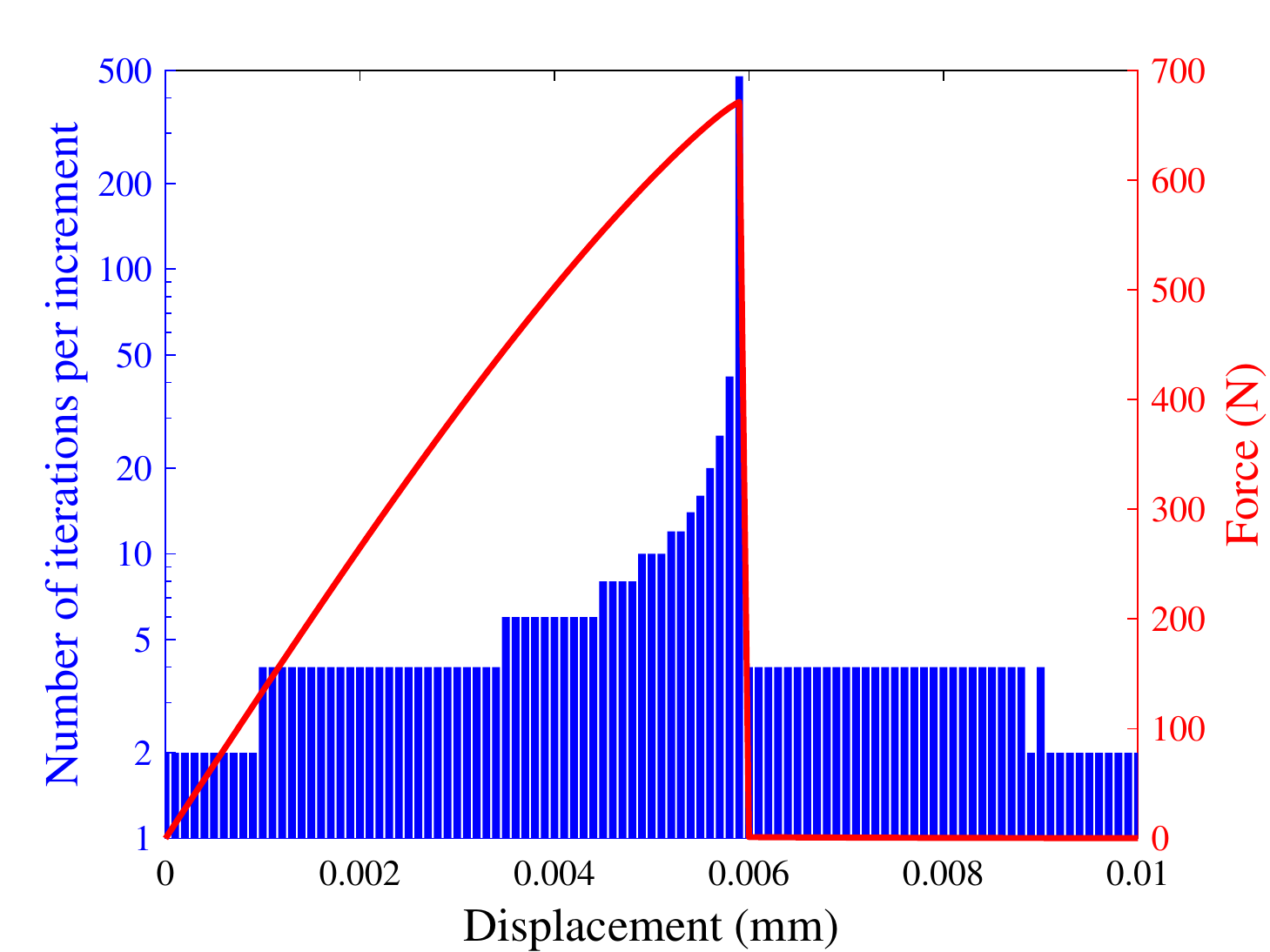}
    \caption{Notched square plate under tension. Number of iterations per increment, with the force versus displacement curve superimposed.}
    \label{Fig:Sq-for}
\end{figure}

The force versus displacement response predicted is shown in Fig. \ref{Fig:Sq-for}. The result agrees with that of \citet{TAFM2020}, which was obtained using a quasi-Newton solution scheme. Cracking is unstable, with the crack extending through the ligament instantaneously. This leads to a dramatic drop in the load carrying capacity, as shown in Fig. \ref{Fig:Sq-for}. However, despite this drastic change in the structural response, convergence can be attained and the fracture event is captured in one single load increment. Fig. \ref{Fig:Sq-for} also shows the number of iterations required to achieve convergence in each increment, superimposed to the force versus displacement response. We use time increments of constant size and resolve the analysis with a total of 100 load increments. Convergence throughout can be achieved with as few as 10 increments, but using a larger number facilitates capturing the sudden load drop with greater fidelity. An adaptive time stepping scheme, such as the one developed by \citet{TAFM2020}, can be easily incorporated. This will allow for the increment size to increase or decrease as needed, enabling accurate results at an even smaller computational cost. In any case, it can be observed that the problem can be solved efficiently, with most time increments requiring a small number of iterations to achieve convergence (10 or fewer). However, resolving the fracture event requires a load increment with over 400 iterations. Unlike other computational fracture methods, the Newton-Raphson algorithm can converge after hundreds of iterations in phase field models \citep{Gerasimov2016}. The solution controls of Abaqus have to be edited to increase the maximum number of iterations that are allowed before convergence is deemed unlikely and the load increment is aborted (see the accompanying input file, to be downloaded from www.empaneda.com/codes). It must be noted that, despite the good performance observed, this boundary value problem can be resolved more efficiently using quasi-Newton solution schemes (see \citealp{TAFM2020}).

\subsection{Notched square plate under shear}
\label{Sec:PlateShear}

\begin{figure}[H]
    \centering
    \begin{subfigure}[H]{0.45\textwidth}
        \includegraphics[width=\textwidth]{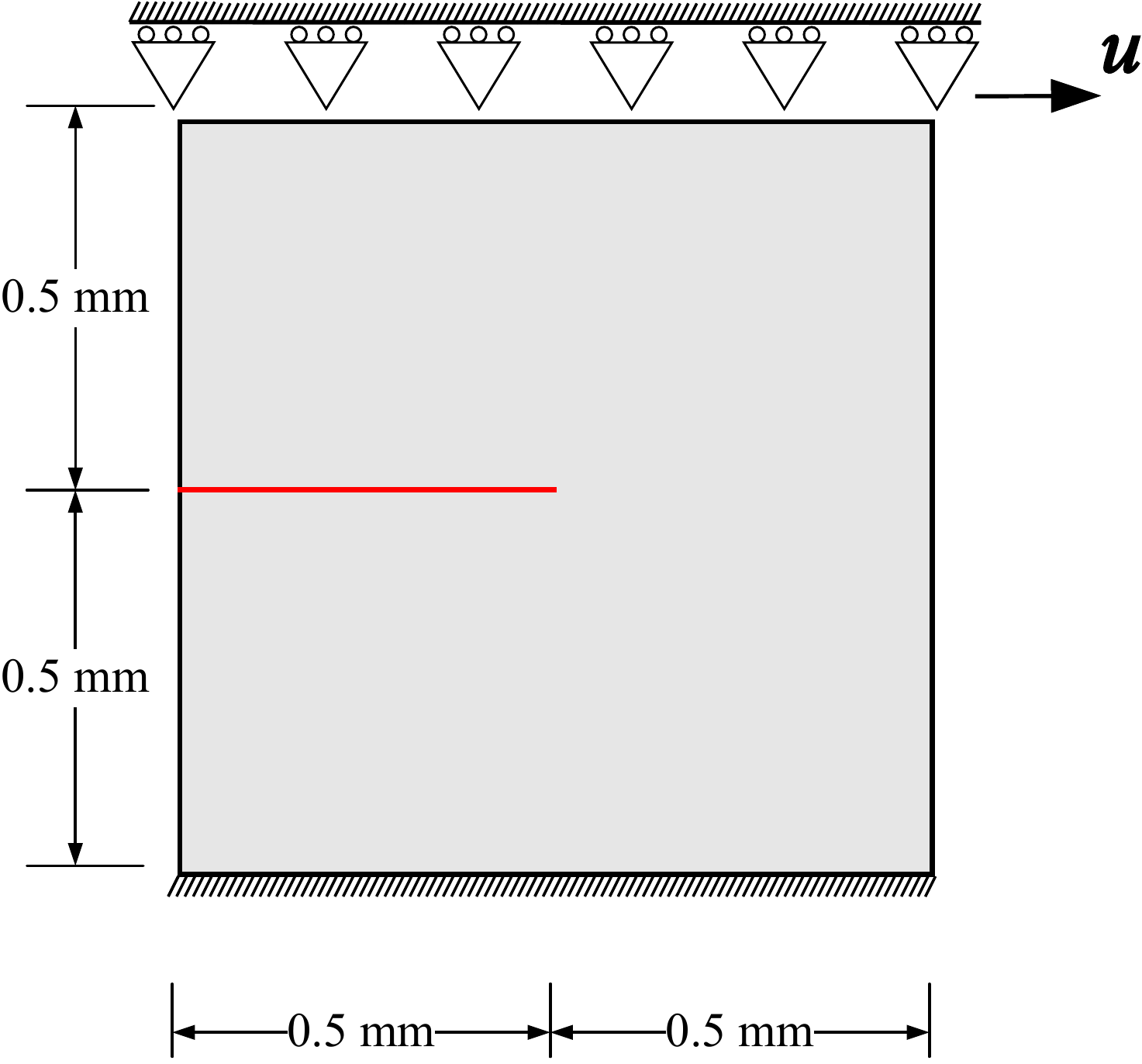}
        \caption{}
        \label{Fig:Sq-sh-geo}
    \end{subfigure}
        \begin{subfigure}[H]{0.45\textwidth}
        \includegraphics[width=\textwidth]{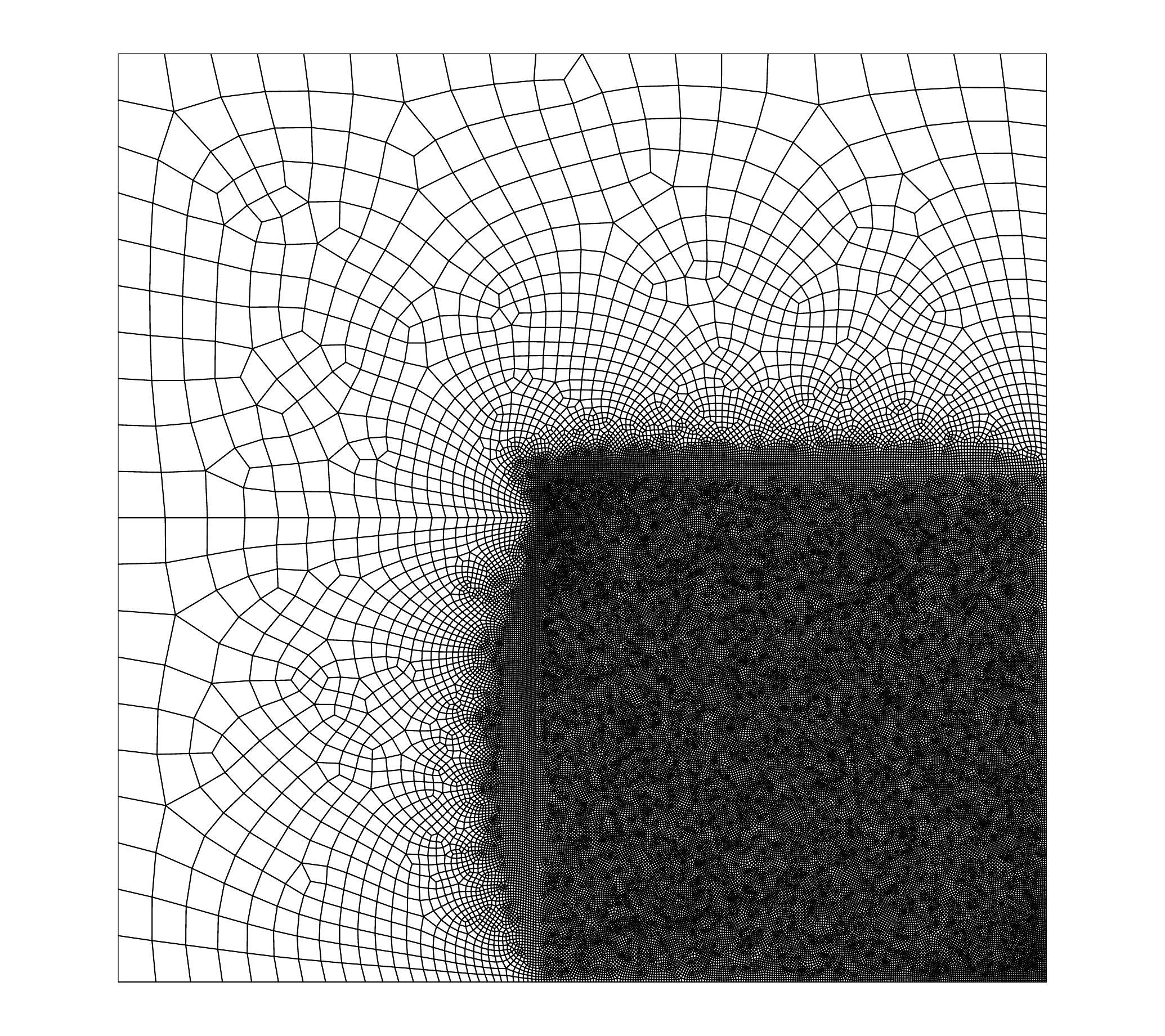}
        \caption{}
        \label{Fig:Sq-sh-mesh}
    \end{subfigure}
    \begin{subfigure}[H]{0.45\textwidth}
        \includegraphics[width=\textwidth]{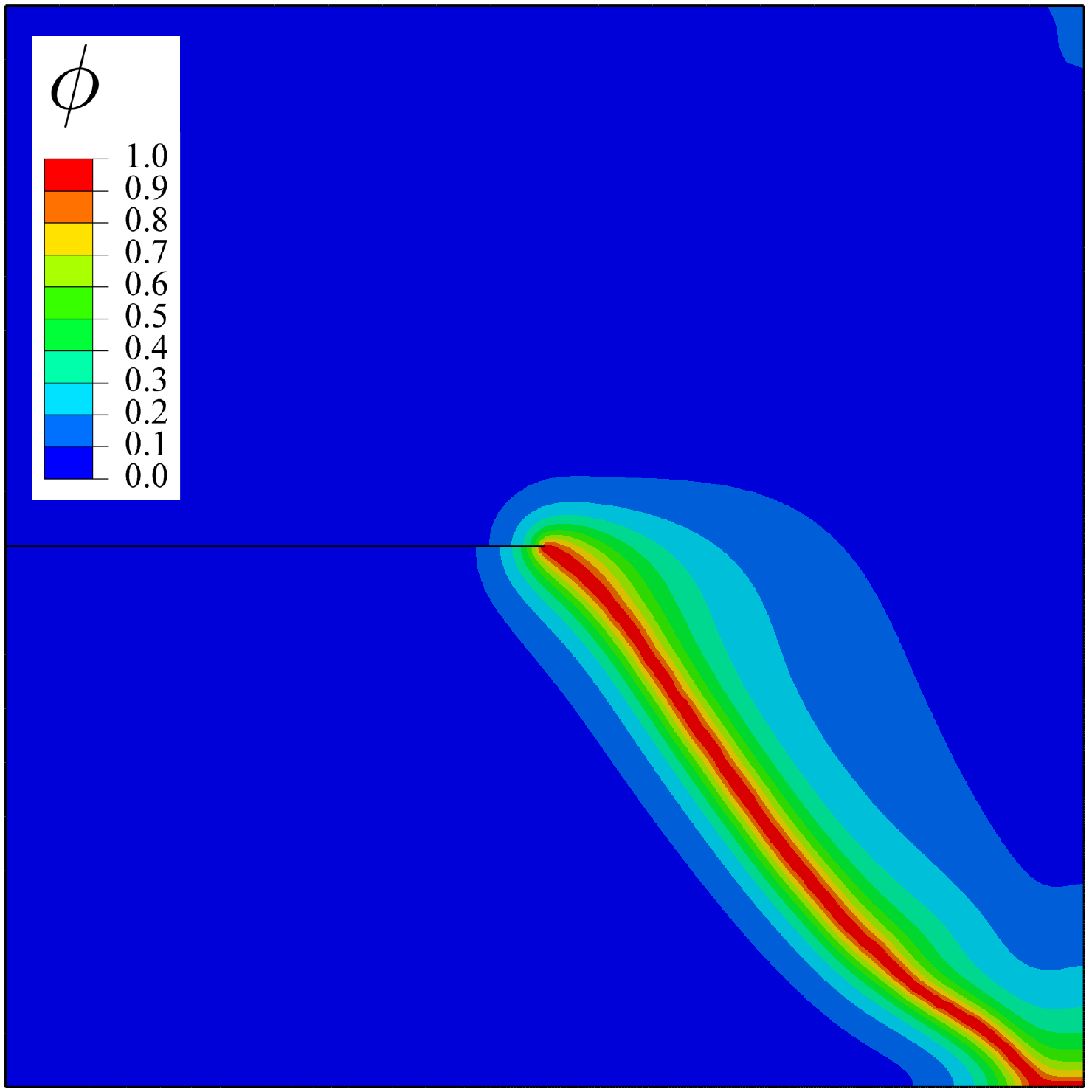}
        \caption{}
        \label{Fig:Sq-sh-phi}
    \end{subfigure}
    \caption{Notched square plate under shear: (a) geometry, dimensions and boundary conditions, (b) finite element mesh, and (c) contour of the phase field $\phi$ after rupture.}
\end{figure}

We shall now address the case of stable crack growth by simulating the fracture of the notched square plate considered in Section \ref{Sec:PlateTension}, but subjected to shear loading. As shown in Fig. \ref{Fig:Sq-sh-geo}, a horizontal displacement is prescribed at the top edge of the plate, while the bottom edge is fully constrained $u_x=u_y=0$. The dimensions of the initial crack and the sample are identical to those considered for the uniaxial tension case study. Also, the same material properties are assumed. On this occasion, the volumetric-deviatoric split of the strain energy density proposed by \citet{Amor2009} is adopted - see \ref{App:FEM}. This is implemented using the so-called hybrid approach by \citet{Ambati2015}, such that the displacement field equation remains as in Eq. (\ref{eqn:strongForm}a). Based on the literature (see, e.g., \citealp{Ambati2015,TAFM2020}), the crack is expected to deflect towards the bottom-right corner. Accordingly, the mesh is refined in the bottom half of the sample - see Fig. \ref{Fig:Sq-sh-mesh}. A total of 73,714 linear quadrilateral elements with full integration are used, with the characteristic element size being ten times smaller than the phase field length scale. The phase field contours at the end of the analysis are provided in Fig. \ref{Fig:Sq-sh-phi}, showing the final crack trajectory. The crack path predicted agrees with that observed in previous studies using the volumetric-deviatoric split \citep{Ambati2015,TAFM2020}.

\begin{figure}[H]
    \centering
    \begin{subfigure}[H]{0.45\textwidth}
    \centering
    \begin{tikzpicture}[node distance=.2cm]
    \node  {Monolithic};
    \end{tikzpicture}
    \includegraphics[width=\textwidth]{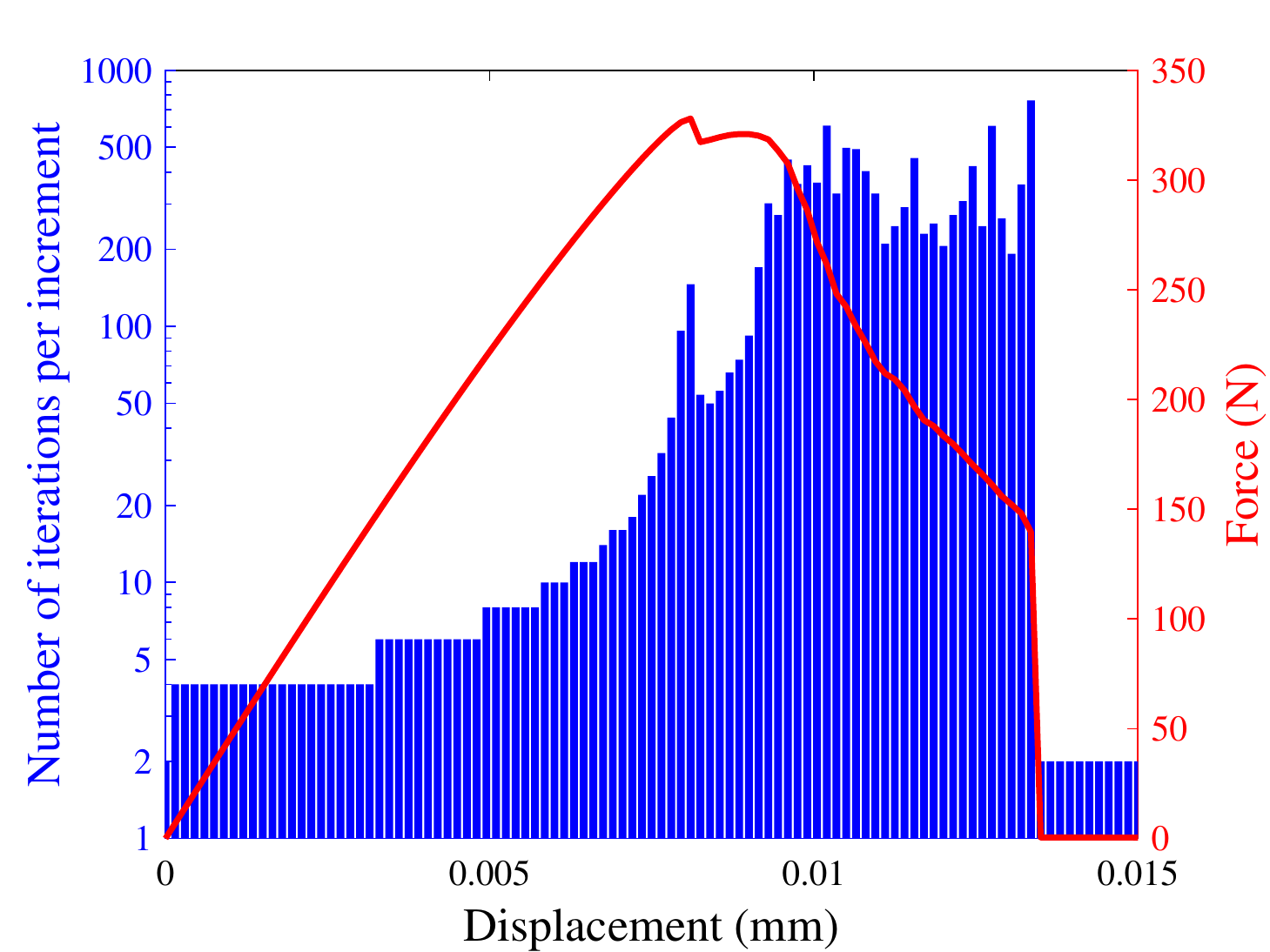}
    \caption{}
    \end{subfigure}\hspace{0.05\textwidth}
    \begin{subfigure}[H]{0.45\textwidth}
        \centering
    \begin{tikzpicture}[node distance=.2cm]
    \node  {Staggered};
    \end{tikzpicture}
    \includegraphics[width=\textwidth]{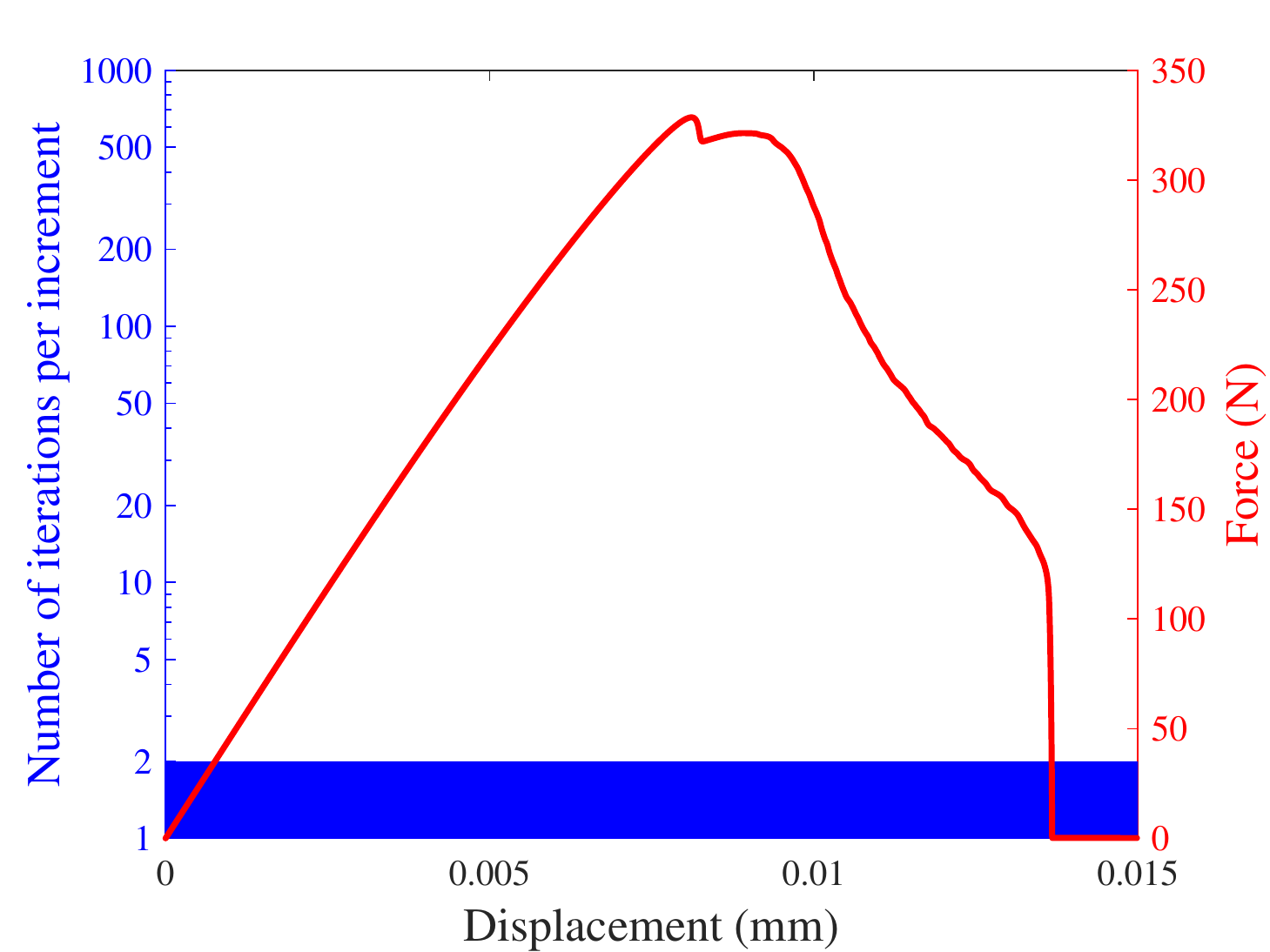}
    \caption{}
    \end{subfigure}
    \begin{subfigure}[H]{0.6\textwidth}
    \centering
    \includegraphics[width=\textwidth]{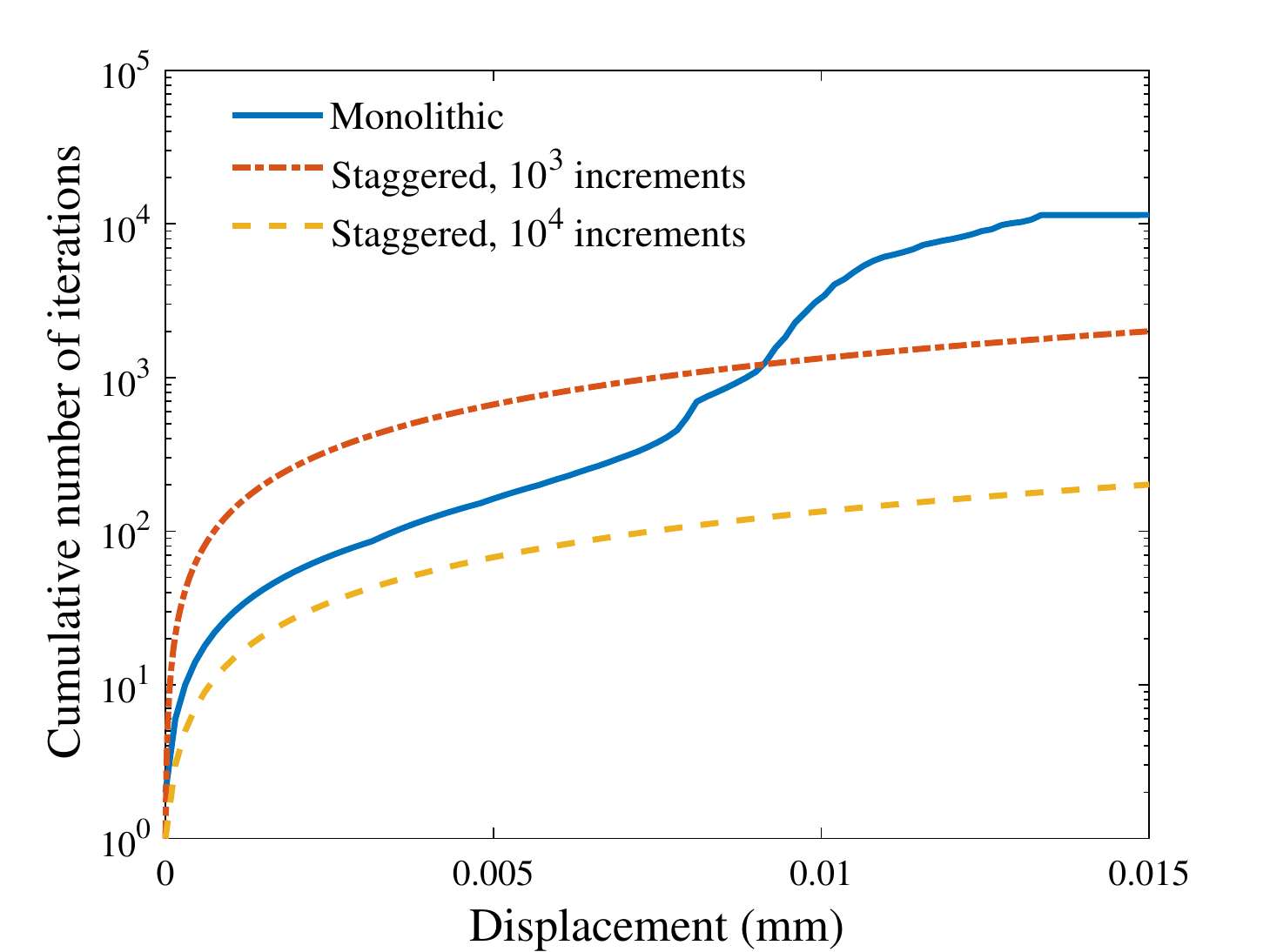}
    \caption{}
    \end{subfigure}
    \caption{Notched  square  plate  under  shear: (a) Number of iterations per increment for the monolithic scheme, with the force versus displacement curve superimposed; (b) number of iterations per increment for the staggered scheme with 10,000 increments, with the force versus displacement curve superimposed; and (c) cumulative number of iterations for both staggered and monolithic results.}
    \label{Fig:Sc-sh-res}
\end{figure}
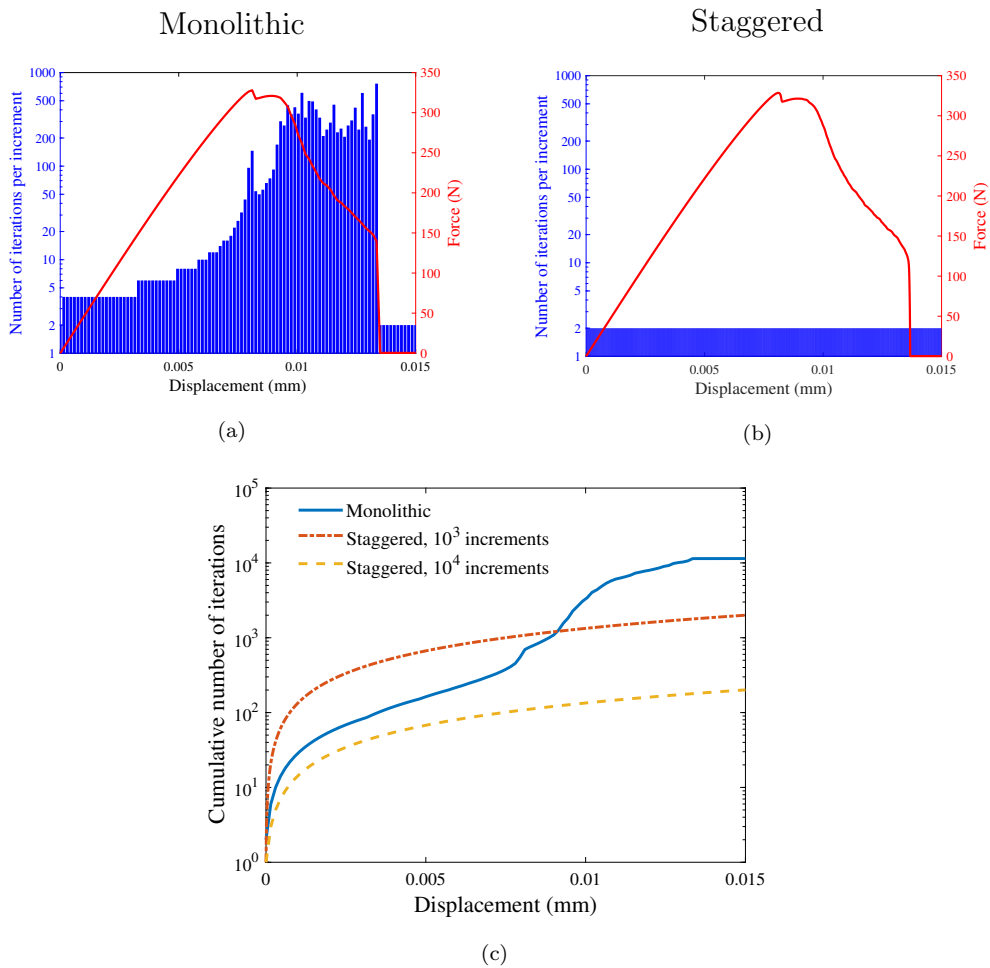

The force versus displacement response is shown in Fig. \ref{Fig:Sc-sh-res}, along with the size of each increment and the number of iterations that were needed to achieve convergence. The crack propagates in a stable manner, leading to a progressive reduction in the reaction force. Again, the results agree with those obtained by \citet{TAFM2020} using a monolithic quasi-Newton solution scheme. This boundary value problem is known to be particularly challenging from a convergence viewpoint and is thus used to compare the monolithic and staggered solution schemes. Consider first the monolithic analysis, Fig. \ref{Fig:Sc-sh-res}a. While the entire crack propagation process can be captured, many increments require a very significant number of iterations to achieve convergence - unlike in the uniaxial tension case where cracking is unstable. It is clear that, for this boundary value problem, the monolithic implementation struggles to converge and becomes inefficient. Now let us examine the output of the staggered case. The results obtained with the single-pass staggered implementation also make use of a uniform increment size, with the entire analysis being conducted using $10^4$ load steps. This is a sufficiently large number of increments such that the solution is similar to that obtained with the unconditionally stable monolithic model - see Fig. \ref{Fig:Sc-sh-res}b. In the staggered case, all load steps converge after two increments. Notwithstanding, as discussed before, this solution scheme is not unconditionally stable and results can be sensitive to the number of time increments. We also conduct the analysis using $10^3$ load steps; the crack trajectory and the maximum force attained agree with those predicted with the monolithic scheme but the force versus displacement result differs in the softening region (not shown). The staggered implementation appears to be more robust and efficient than the monolithic one for this specific case study; as quantified in Fig. \ref{Fig:Sc-sh-res}c, the total number of iterations is larger in the monolithic case. However, one should note that both implementations are significantly outperformed by a monolithic approach based on the quasi-Newton solution method. As shown in \citep{TAFM2020}, a precise solution to this specific boundary value problem can be obtained with a number of iterations that is one order of magnitude smaller than the accurate staggered solution. 

\subsection{Screw tension tests}
\label{Sec:Screw}

\begin{figure}[H]
    \centering
    \includegraphics[width=0.5\textwidth]{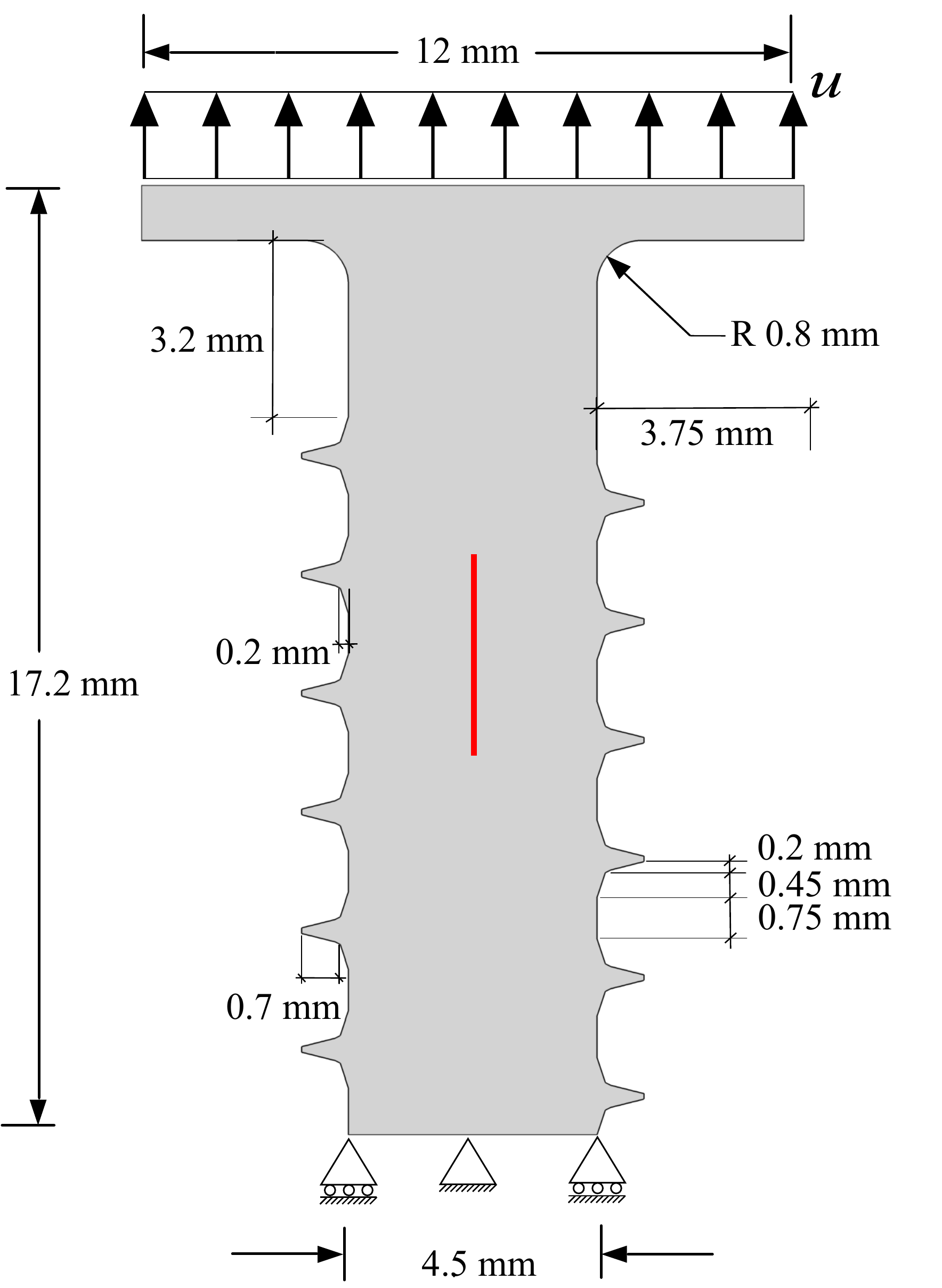}
    \caption{Screw tension tests: geometry, dimensions and boundary conditions.}
    \label{Fig:Sc-geo}
\end{figure}

We proceed now to simulate the fracture of a screw subjected to tension, following the work by \citet{Wick2015}. The geometry, dimensions and boundary conditions mimic those of \citep{Wick2015} and are shown in Fig. \ref{Fig:Sc-geo}. Three different cases are considered. First, we model a screw with no initial damage; i.e., without the initial crack displayed in Fig. \ref{Fig:Sc-geo}. Secondly, we will assume that the screw contains an initial short crack, with size $a=3$ mm. Thirdly, a screw with a long crack will be modelled, where $a=6$ mm. In all cases, the initial cracks are introduced by defining as initial condition $\phi=1$. Moreover, the initial crack is vertical, as shown in Fig. \ref{Fig:Sc-geo}, has a thickness of 0.16 mm, and its bottom tip is located at a distance of 7 mm to the bottom of the screw. Following \citet{Wick2015}, the material properties are taken to be $E = 210$ GPa, $\nu = 0.3$, $\ell = 0.2$ mm, and $G_c = 2.7$ N/mm. The screws are discretised using approximately 70,000 linear quadrilateral elements. The samples are meshed uniformly so as to remove any bias of the mesh on the crack trajectory, with the characteristic element size being 5 times smaller than the phase field length scale. Computations are conducted with the monolithic scheme and no strain energy density split is considered.

\begin{figure}[H]
    \centering
        \begin{subfigure}[H]{0.07\textwidth}
        \begin{tikzpicture}[node distance=2cm]
    \node (cap) {{\Large $\phi$}};
    \end{tikzpicture}
        \includegraphics[width=\textwidth]{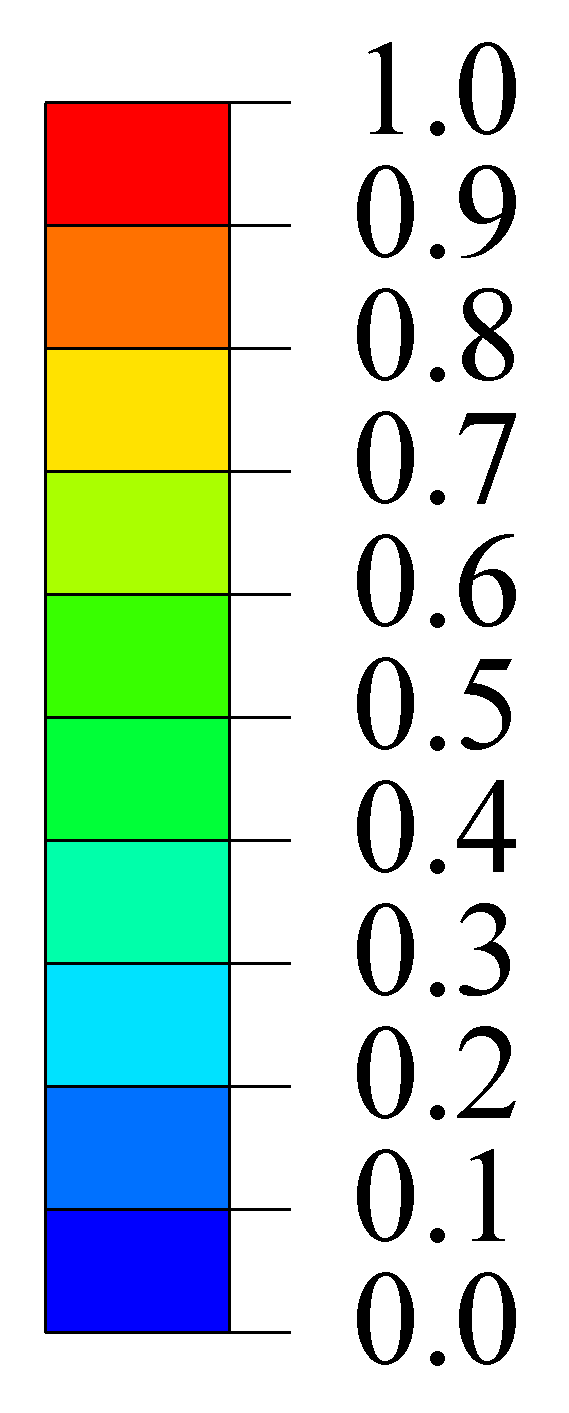}
    \end{subfigure}
    \begin{subfigure}[H]{0.3\textwidth}
        \includegraphics[width=\textwidth]{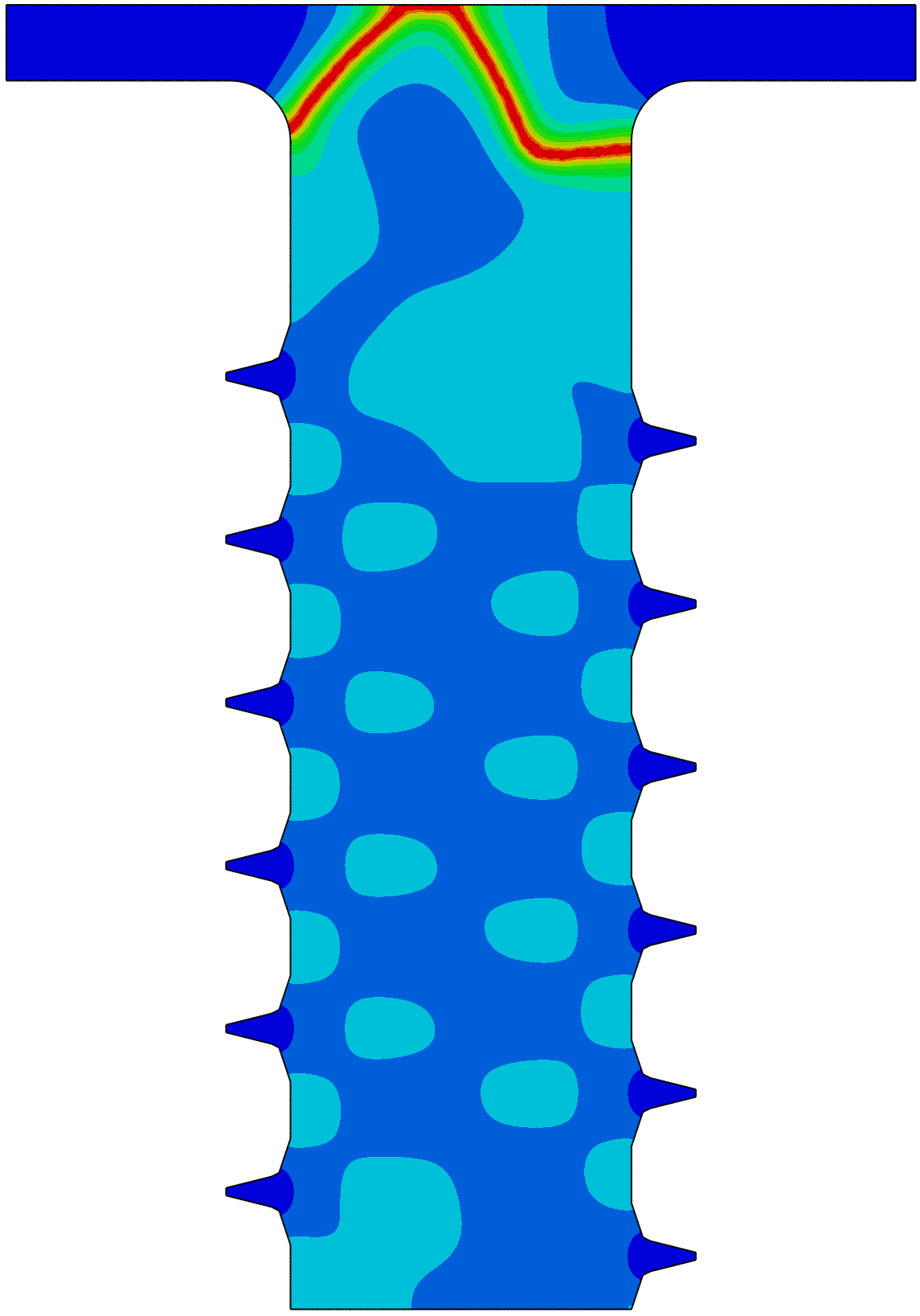}
        \caption{}
    \end{subfigure}
    \begin{subfigure}[H]{0.3\textwidth}
        \includegraphics[width=\textwidth]{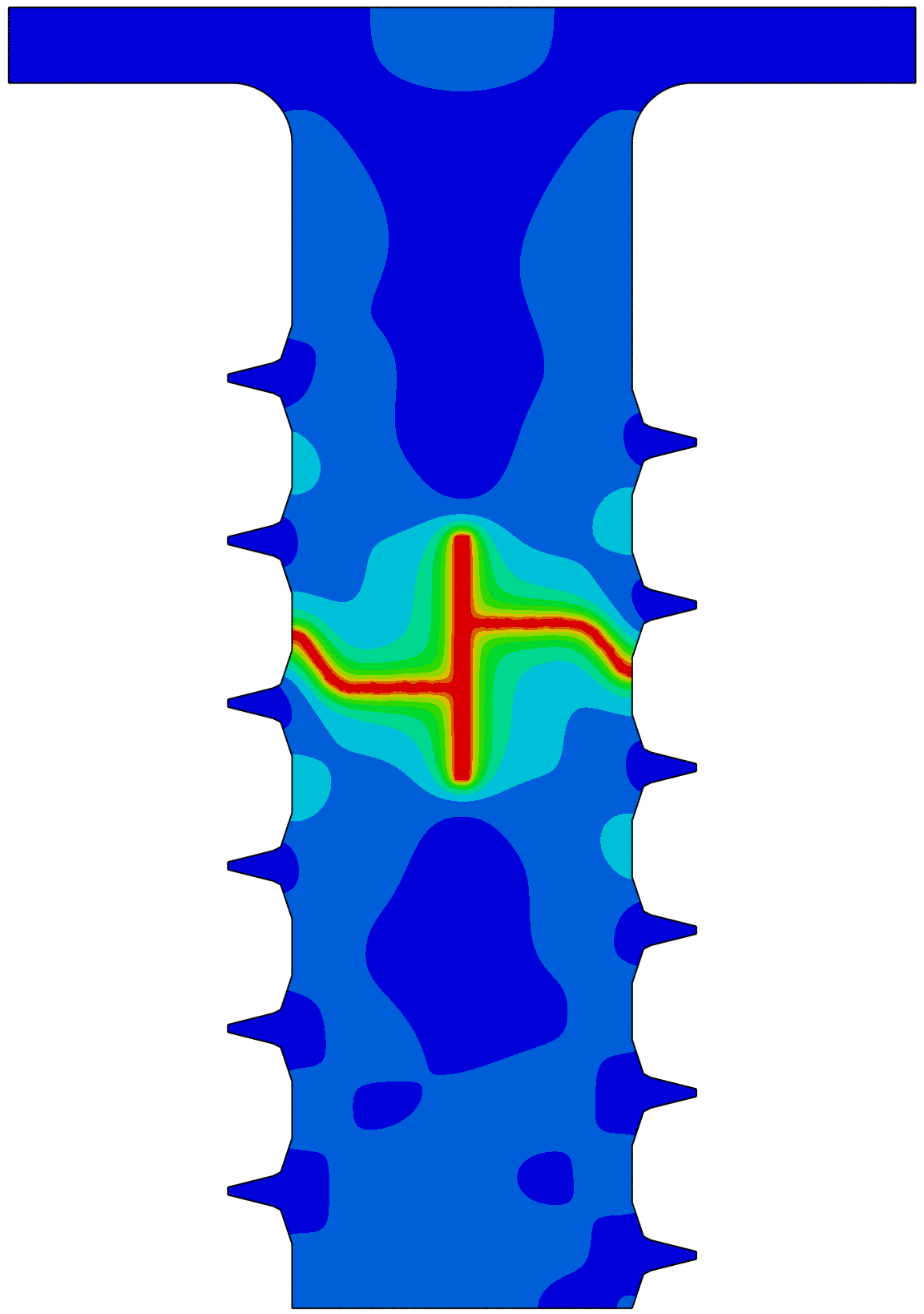}
        \caption{}
    \end{subfigure}
        \begin{subfigure}[H]{0.3\textwidth}
        \includegraphics[width=\textwidth]{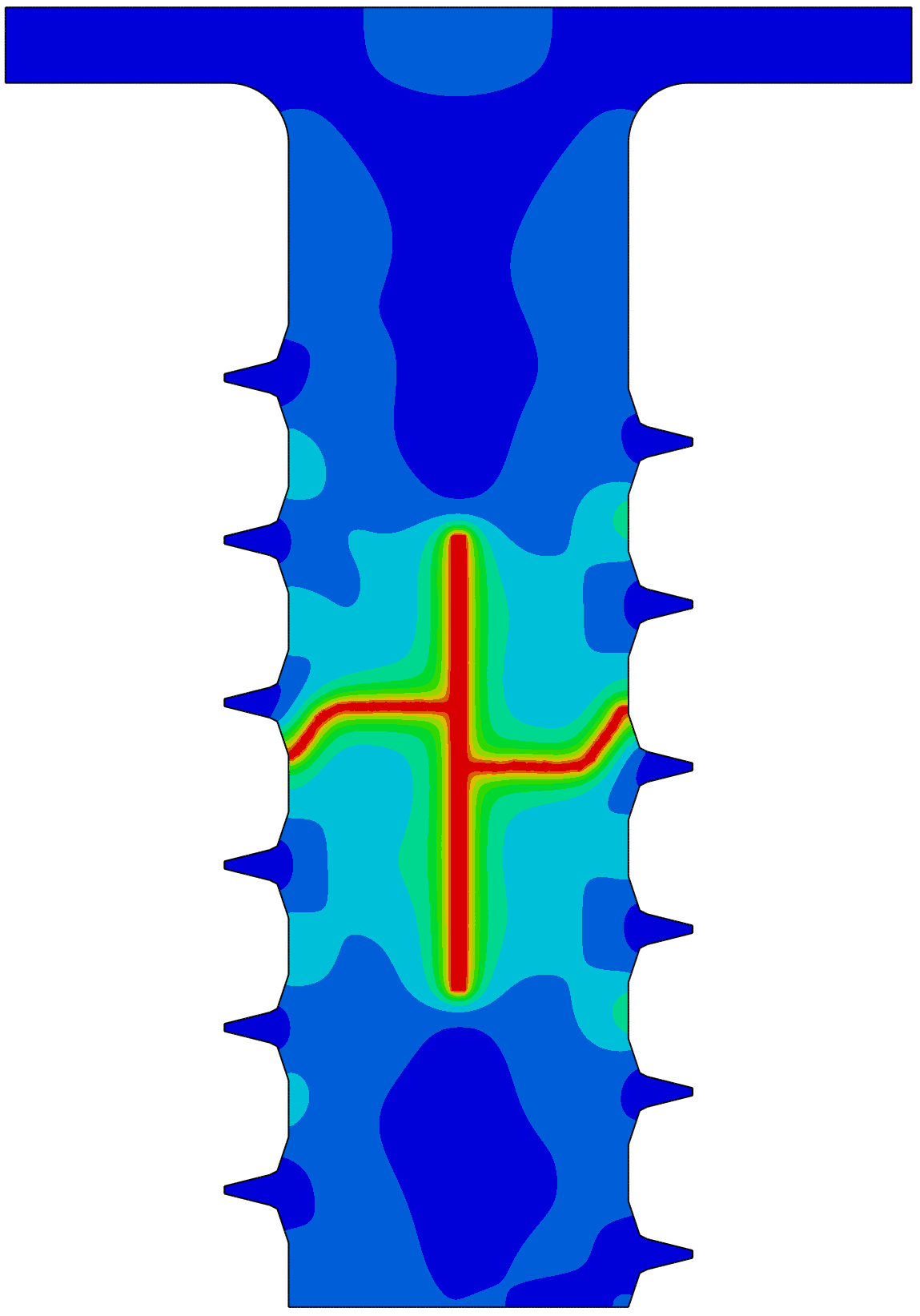}
        \caption{}
    \end{subfigure}
    \caption{Screw tension tests: final phase field contours for the cases of (a) a screw with no initial crack, (b) a screw with a short ($a=3$ mm) initial crack, and (c) a screw with a long ($a=6$ mm) initial crack.}
    \label{Fig:Sc-phi}
\end{figure}
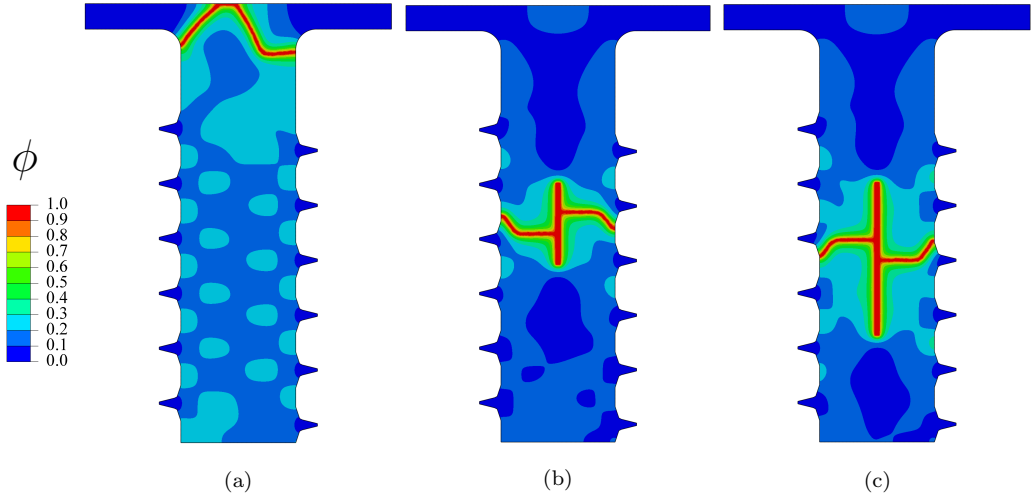

The crack growth trajectories predicted for the three cases described above are shown in Fig. \ref{Fig:Sc-phi}, by plotting the phase field contours. The results agree qualitatively with those obtained by \citet{Wick2015}. In the absence of an initial defect, crack nucleation takes place near the head of the screw. This is in agreement with expectations, as the first winding of the thread carries the highest load (see \citealp{TAFM2021}). However, when an initial defect is present, two cracks branch from it and propagate until reaching the sides of the screw.

\begin{figure}[H]
    \centering
    \begin{subfigure}[H]{0.48\textwidth}
        \includegraphics[width=\textwidth]{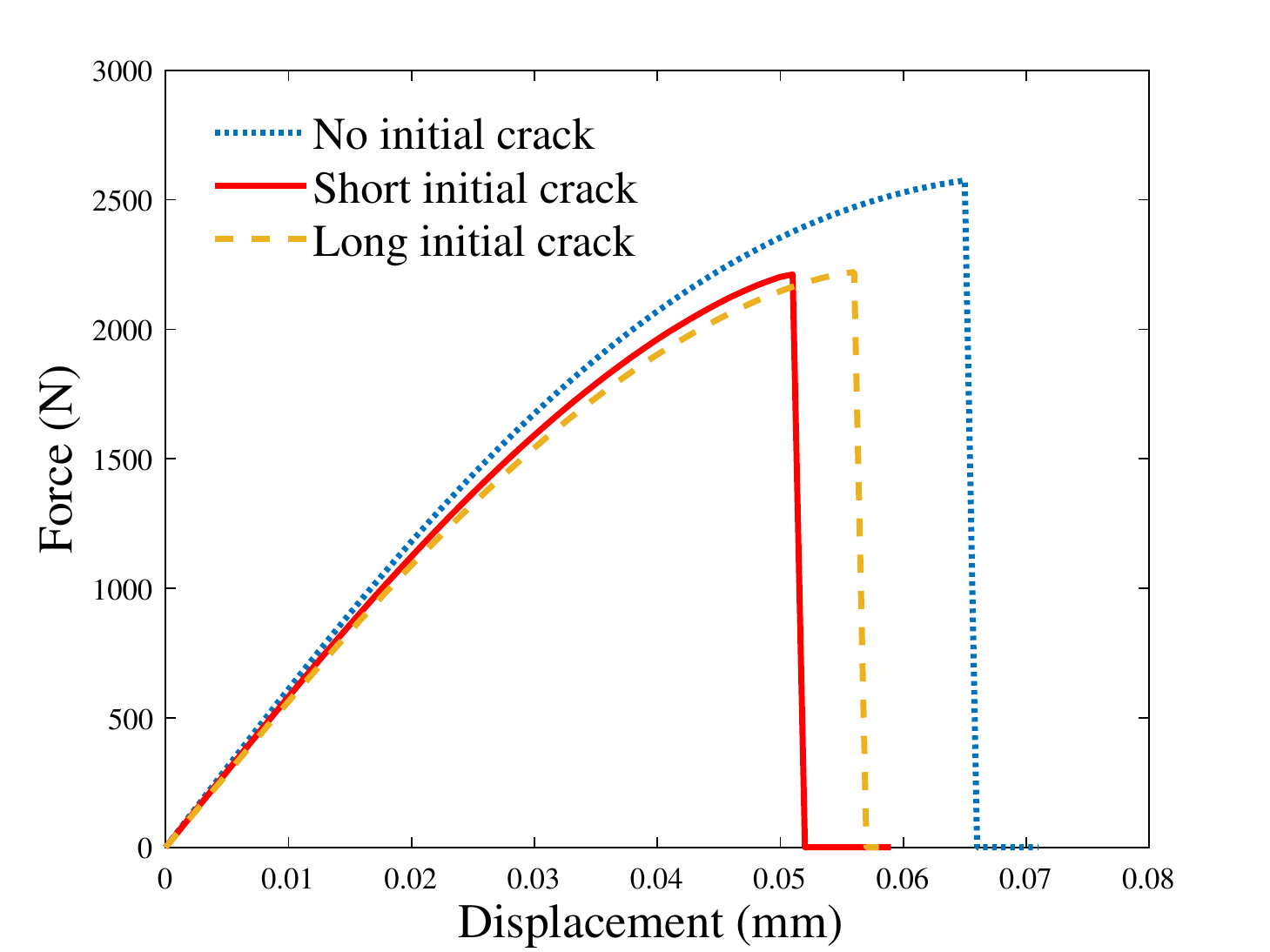}
        \caption{}
        \label{Fig:Sc-for-a}
    \end{subfigure}
    \begin{subfigure}[H]{0.48\textwidth}
        \includegraphics[width=\textwidth]{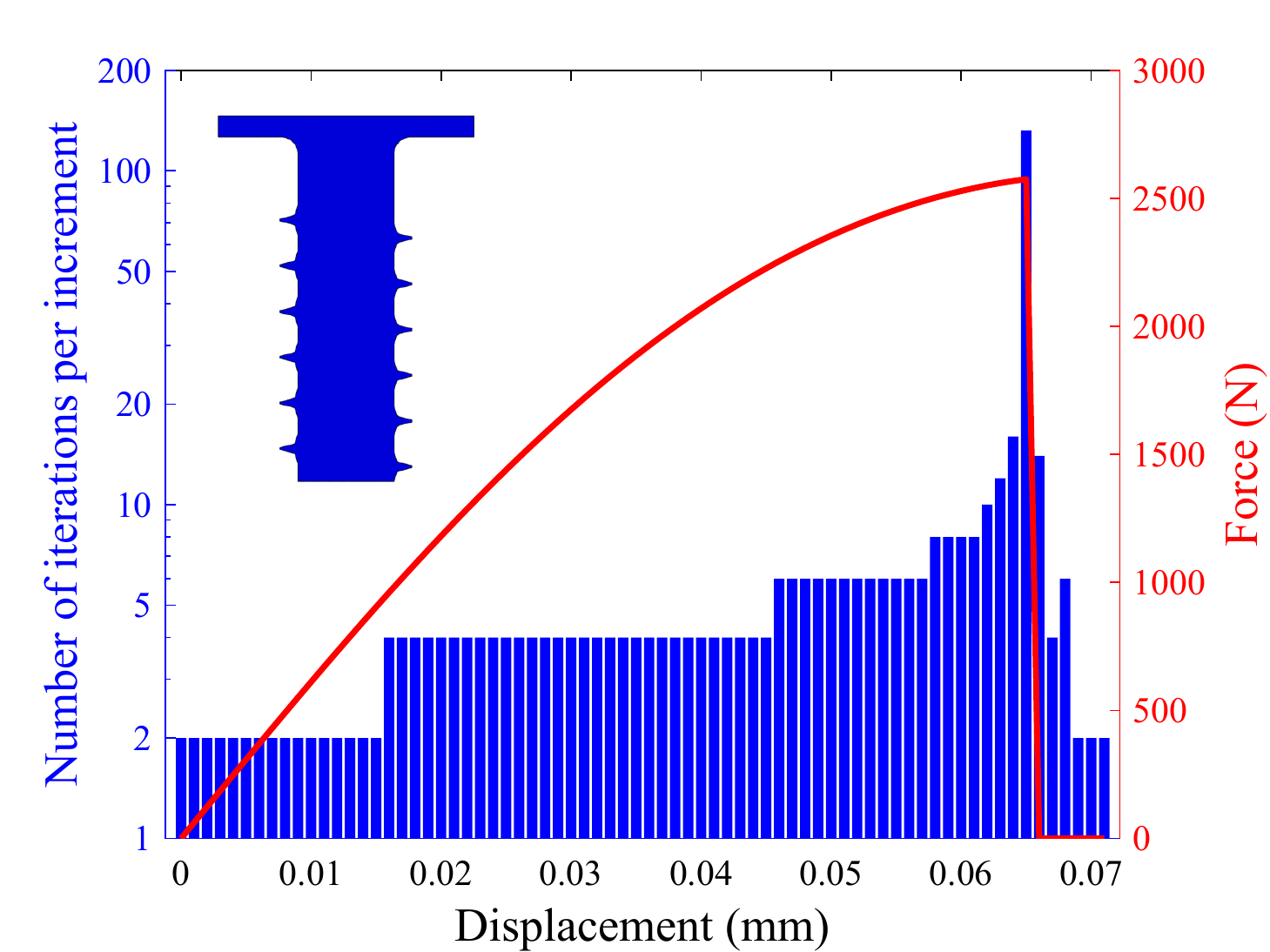}
        \caption{}
        \label{Fig:Sc-for-b}
    \end{subfigure}
        \begin{subfigure}[H]{0.48\textwidth}
        \includegraphics[width=\textwidth]{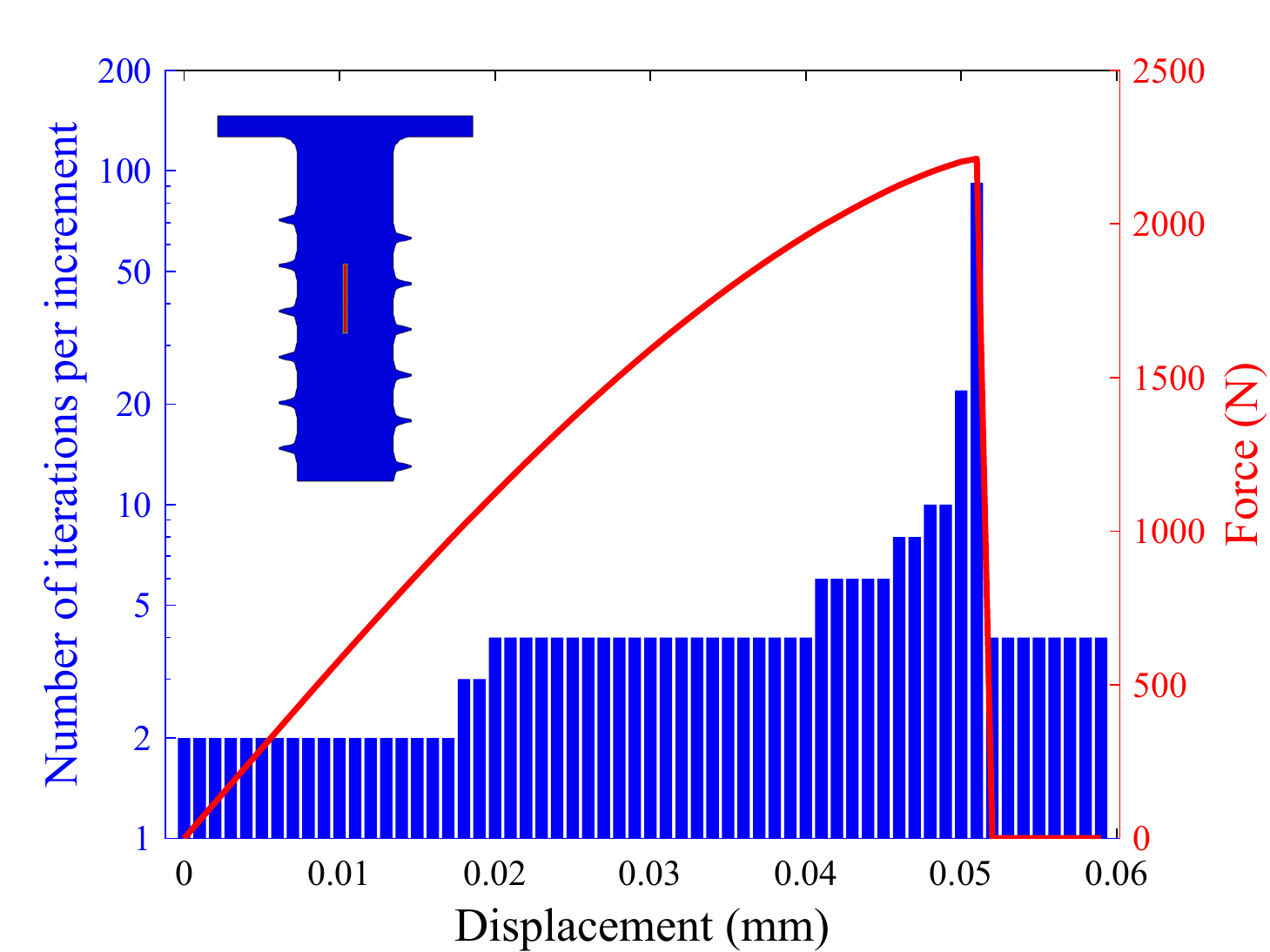}
        \caption{}
        \label{Fig:Sc-for-c}
    \end{subfigure}
    \begin{subfigure}[H]{0.48\textwidth}
        \includegraphics[width=\textwidth]{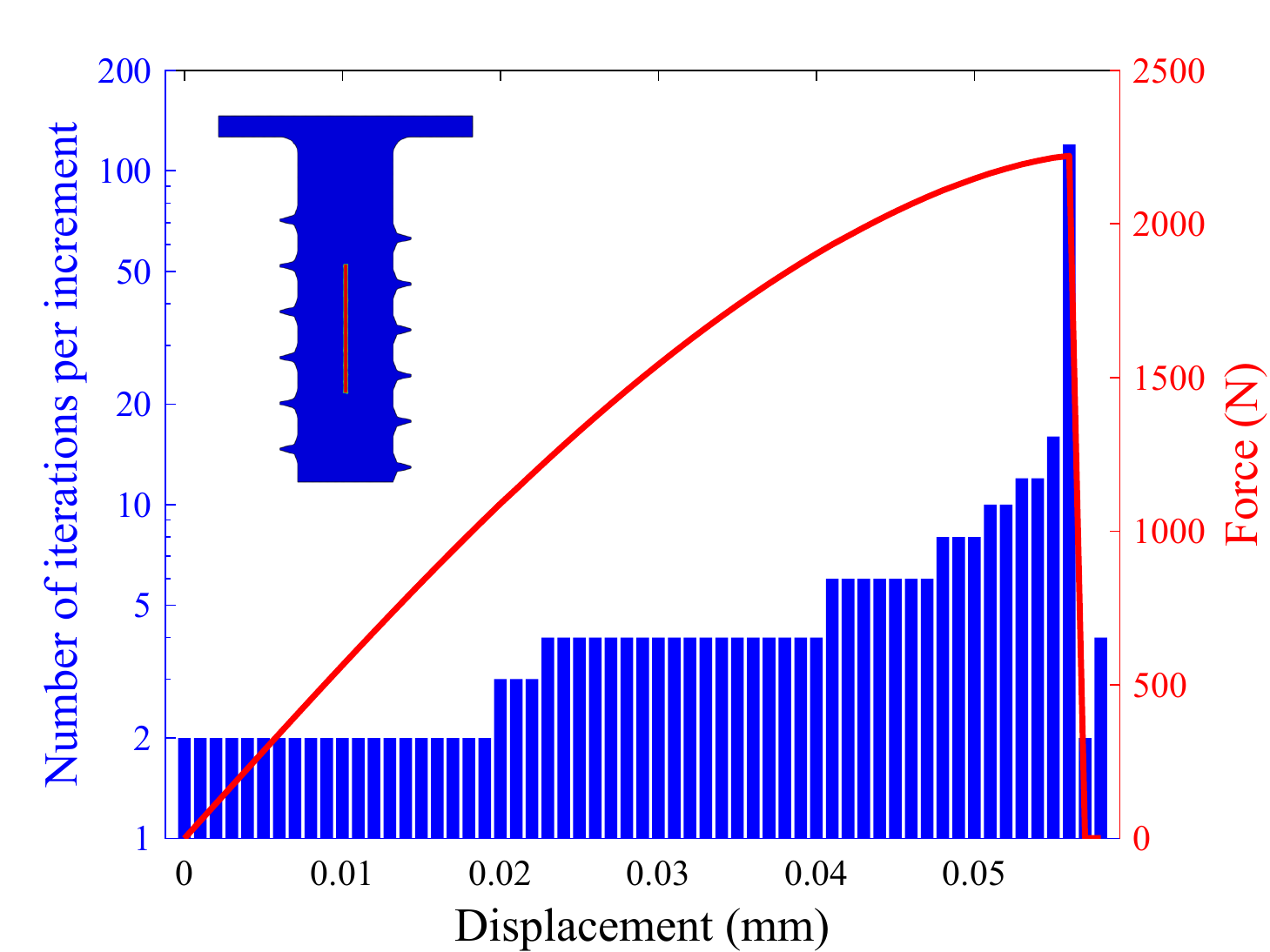}
        \caption{}
        \label{Fig:Sc-for-d}
    \end{subfigure}
    \caption{Screw tension tests: (a) force versus displacement curve for each of the three cases considered, together with the number of iterations per increment for (b) a screw with no initial crack, (c) a screw with a short ($a=3$ mm) initial crack, and (d) a screw with a long ($a=6$ mm) initial crack.}
    \label{Fig:Sc-for}
\end{figure}

The force versus displacement response is shown in Fig. \ref{Fig:Sc-for-a}. In agreement with expectations, the sample without an initial defect is able to carry a larger load. In regard to the screws with an existing defect, the stiffness of the solid is degraded faster in the case of a long crack, relative to the sample with a smaller crack, but the magnitude of the maximum force attained is similar in both cases. The number of iterations required to achieve convergence is shown for every load increment in Figs. \ref{Fig:Sc-for-b}-\ref{Fig:Sc-for-d} for, respectively, the case without an initial defect, the case with an initial long crack and the case with an initial short crack. In all three cases convergence can be readily attained. The crack grows in an unstable fashion and the situation thus resembles that of Section \ref{Sec:PlateTension}; convergence can be readily attained but one specific increment requires more than 100 iterations to do so.

\begin{figure}[H]
    \centering
    \includegraphics[width=1\textwidth]{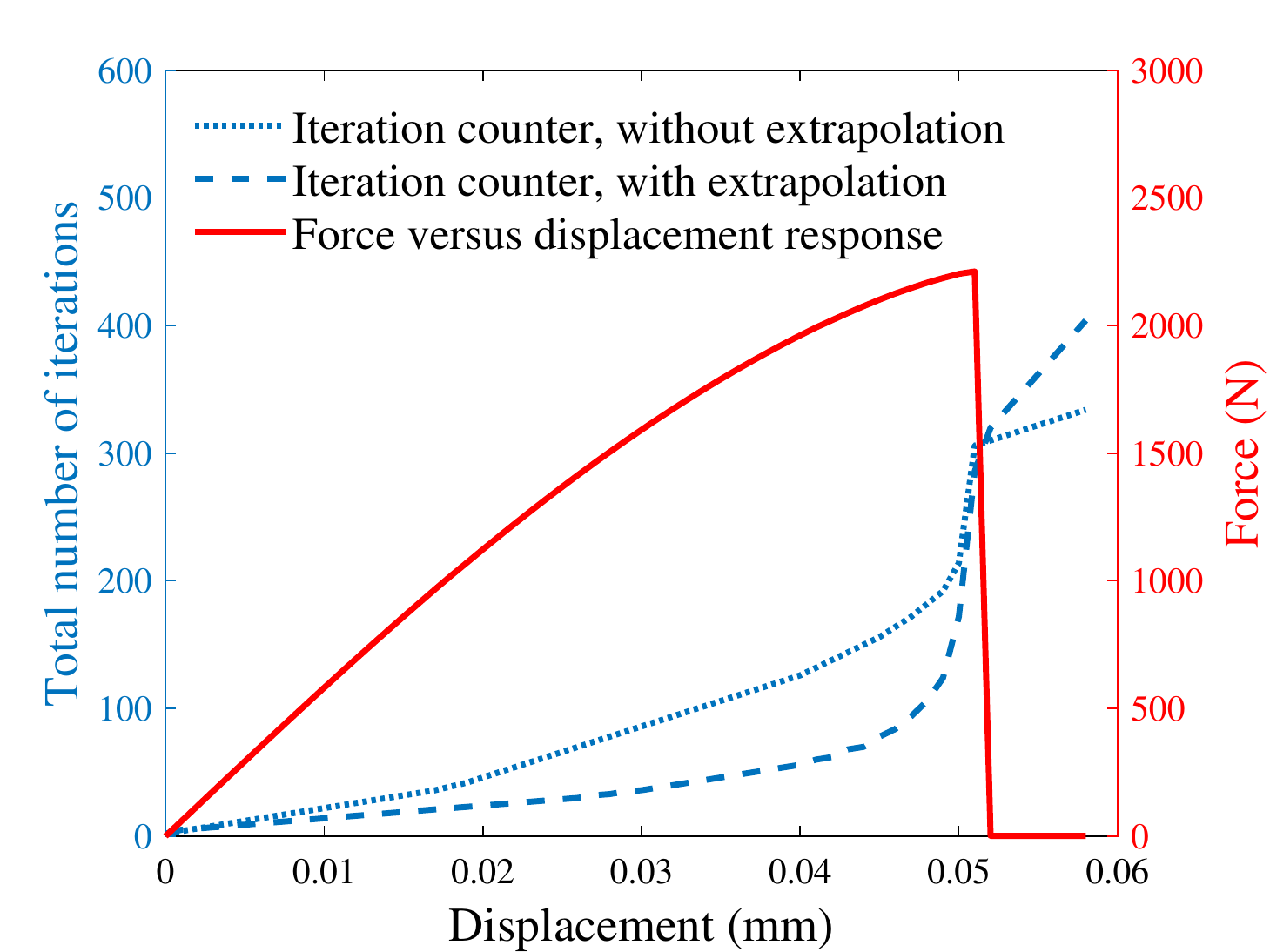}
    \caption{Screw tension test: assessing the influence of the extrapolation technique. Force versus displacement response and cumulative number of iterations required to achieve convergence with and without extrapolation.} 
    \label{Fig:Sc-extrapolation}
\end{figure}

Finally, we investigate the role of using extrapolation to speed up the solution. By default, Abaqus uses linear extrapolation to determine the first guess of the incremental solution. Fig. \ref{Fig:Sc-extrapolation} shows the accumulated number of iterations for the case of a screw with a short initial defect, as a function of the applied displacement and with the force versus displacement response superimposed. It can be readily seen that enabling extrapolation facilitates convergence before cracking occurs, but eventually the solution without extrapolation becomes more efficient as it requires less iterations to resolve the crack propagation process. Thus, computational gains might be attained by deactivating the extrapolation option. 

\subsection{3D Brazilian test}
\label{Sec:Brazilian}

\begin{figure}[H]
    \centering
        \begin{subfigure}[H]{0.45\textwidth}
        \includegraphics[width=\textwidth]{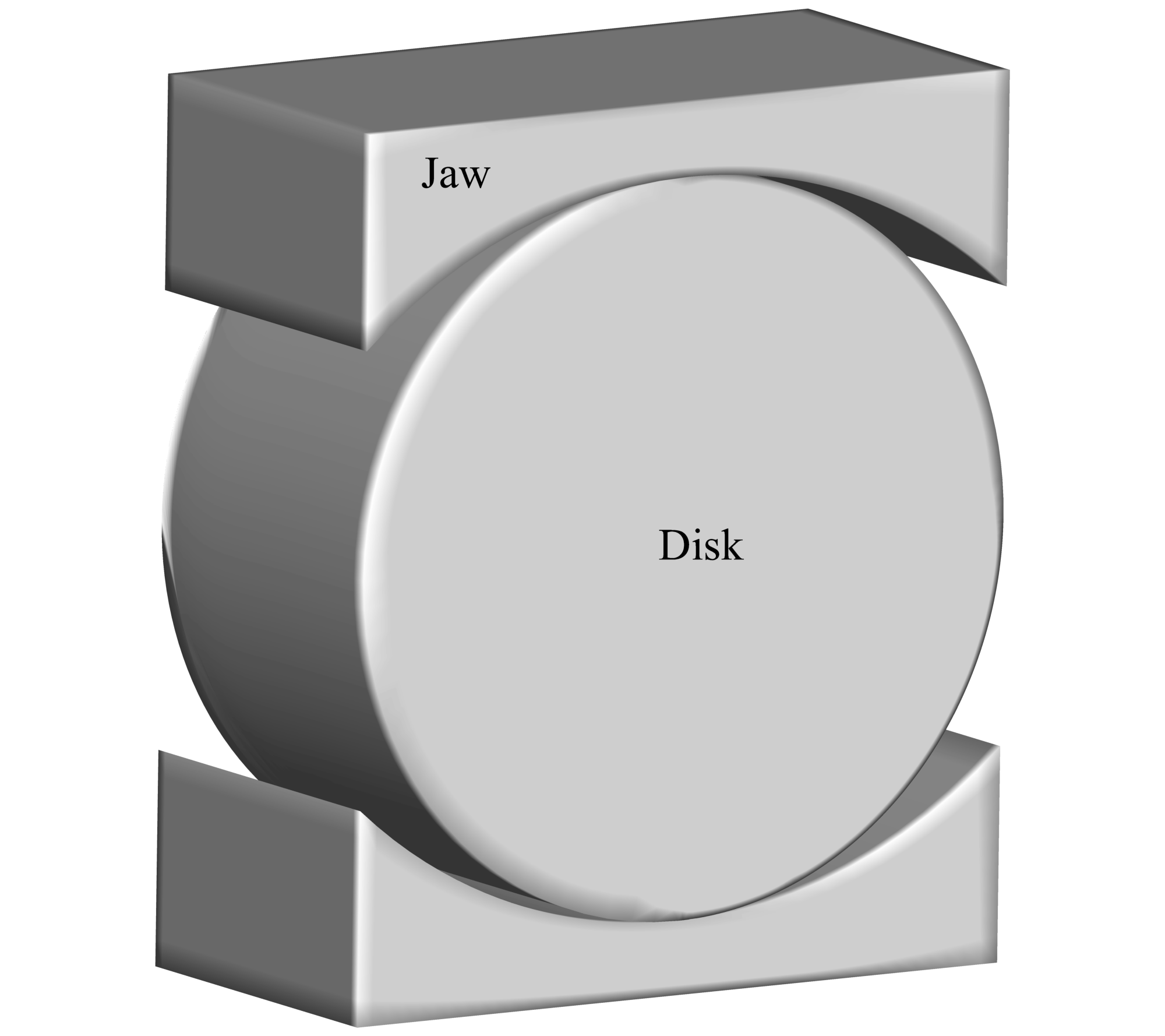}
        \caption{}
        \label{Fig:BT-geo}
    \end{subfigure}
    \begin{subfigure}[H]{0.45\textwidth}
        \includegraphics[width=\textwidth]{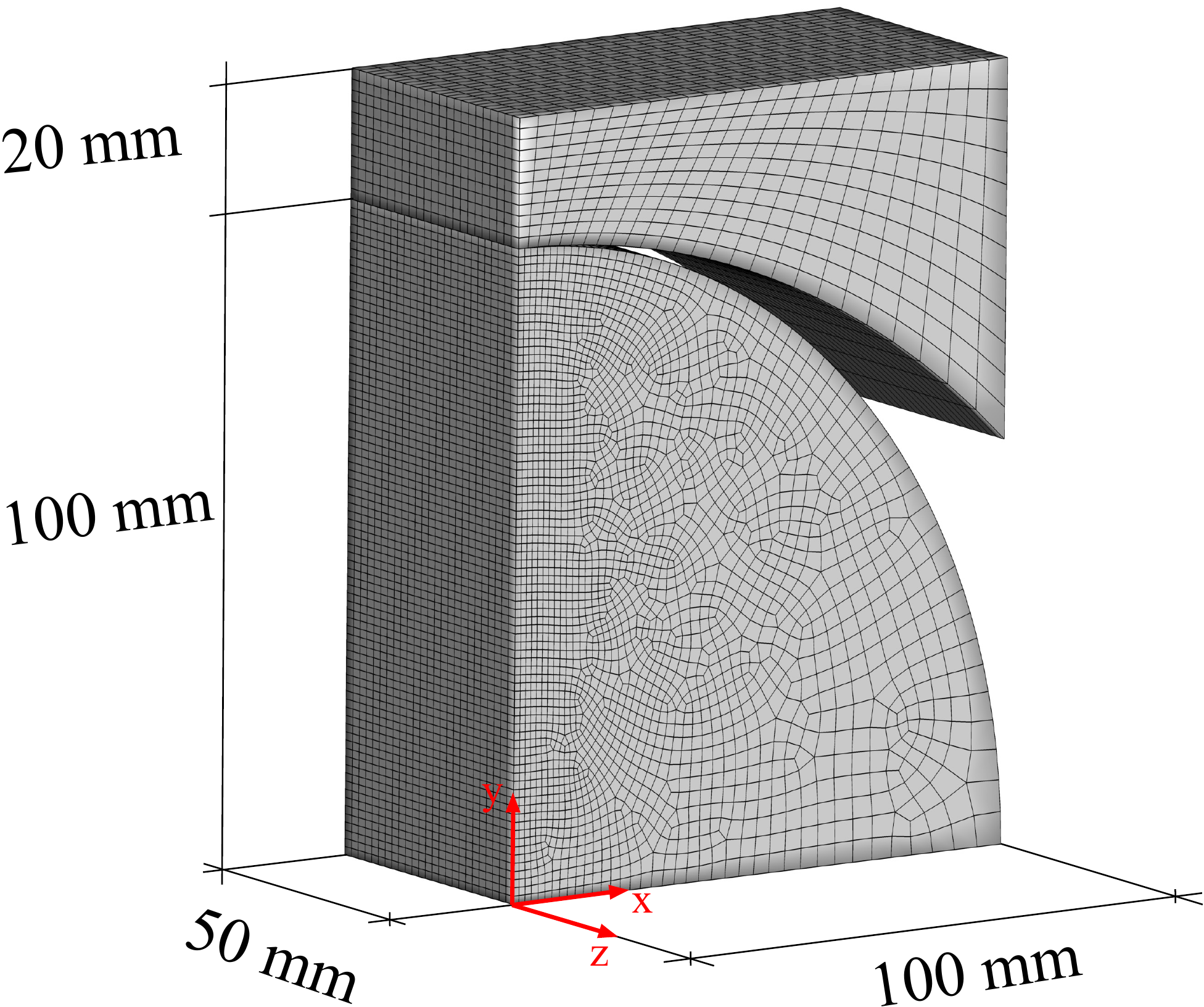}
        \caption{}
        \label{Fig:BT-mesh}
    \end{subfigure}\hspace{0.1\textwidth}
    \caption{3D Brazilian test: (a) complete geometry of the test and (b) geometry, boundary conditions and mesh of the computational model. One-eighth of the problem is simulated, taking advantage of symmetry.}
\end{figure}

Finally, we showcase the potential of the framework presented in capturing structural failure in 3D solids. We do so by simulating the fracture of a brittle solid subjected to the Brazilian test. The Brazilian test is a laboratory experiment widely used in the rock mechanics community to indirectly measure the tensile strength of brittle materials. As shown in Fig. \ref{Fig:BT-geo}, a circular disk is compressed between two jaws until fracture occurs. Upon the assumption that failure occurs at the centre of the disk, closed form expressions can be used to determine the material tensile strength from the remote load \citep{Garcia-Fernandez2018}. As shown in Fig. \ref{Fig:BT-mesh}, we take advantage of symmetry and model one-eighth of the experiment applying suitable boundary conditions. Thus, we prescribe $u_z=0$ in the $xy$ plane at $z=0$ for both the disk and the jaw. To account for symmetry about a plane with $x$=constant, we prescribe $u_x=0$ along the $yz$ plane at $x=0$ on the surfaces of the disk and the jaw. Finally, to account for symmetry along the $y$ axis, we constrain $u_y=0$ on the bottom surface of the disk. The compressive load state is achieved by prescribing a negative $u_y$ displacement on the nodes located on the top surface of the jaw. This one-eighth part of the complete testing configuration is discretised using 58,925 linear brick elements. The characteristic element length equals 0.1 mm and the calculations involved 254,384 degrees-of-freedom.

The material properties are defined as follows. On the one side, the jaws are typically made of steel, for which $E=210$ GPa and $\nu=0.3$ are assumed. For the disk we consider a brittle solid with elastic properties $E = 25$ GPa and $\nu = 0.2$ and fracture properties $\ell = 0.5$ mm and $G_c = 0.16$ N/mm. The jaws radius to disk radius ratio is chosen to be $R_j/R_d=1.5$. The contact between the jaws and the disk is defined as surface to surface contact with a finite sliding formulation. The normal behaviour is based on a hard contact formulation, where the contact constraint is enforced with a Lagrange multiplier representing the contact pressure in a mixed formulation. The tangential contact behaviour is assumed to be frictionless. To prevent damage under compression, the spectral tension-compression decomposition by \citet{Miehe2010a} is adopted - see \ref{App:FEM}. Also, an \emph{anisotropic} formulation is used, such that the strain energy density split is accounted for in the balance equation for the displacement problem (see \ref{App:FEM} for details).\\

\begin{figure}[H]
    \centering
    \begin{subfigure}[H]{0.4\textwidth}
        \includegraphics[width=\textwidth]{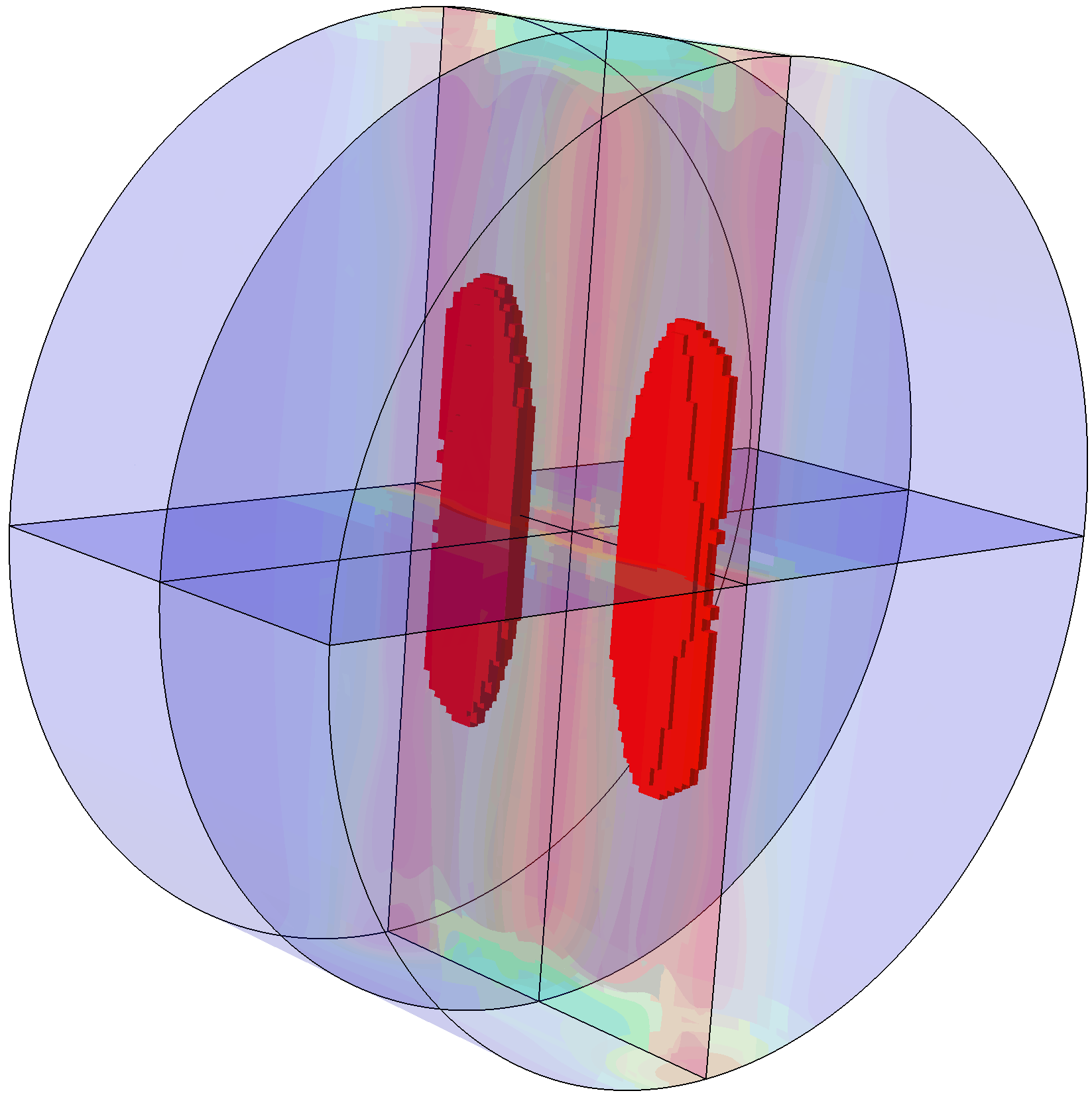}
        \caption{}
    \end{subfigure}\hspace{0.1\textwidth}
    \begin{subfigure}[H]{0.4\textwidth}
        \includegraphics[width=\textwidth]{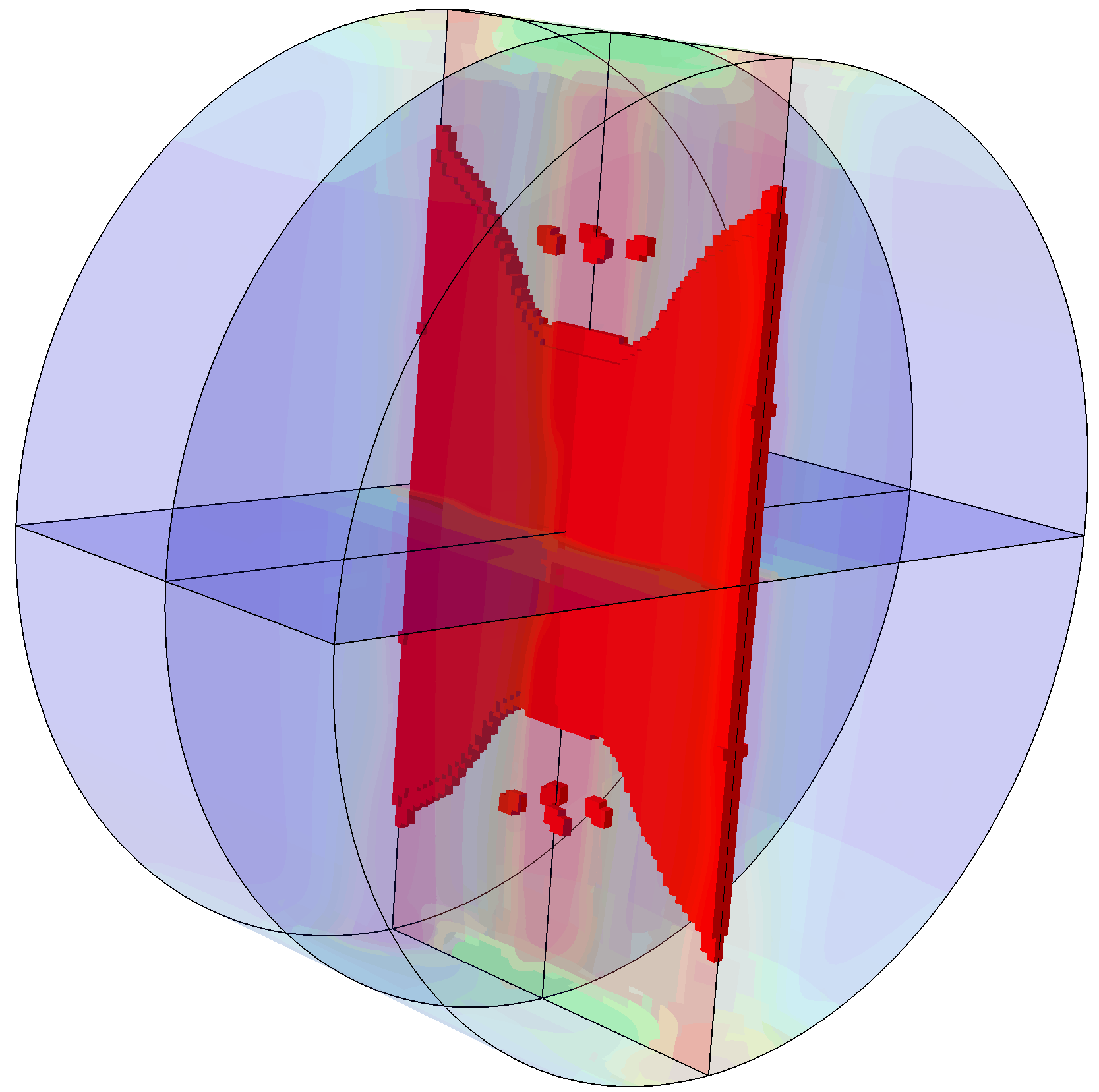}
        \caption{}
    \end{subfigure}\vspace{0.05\textwidth}
        \begin{subfigure}[H]{0.4\textwidth}
        \includegraphics[width=\textwidth]{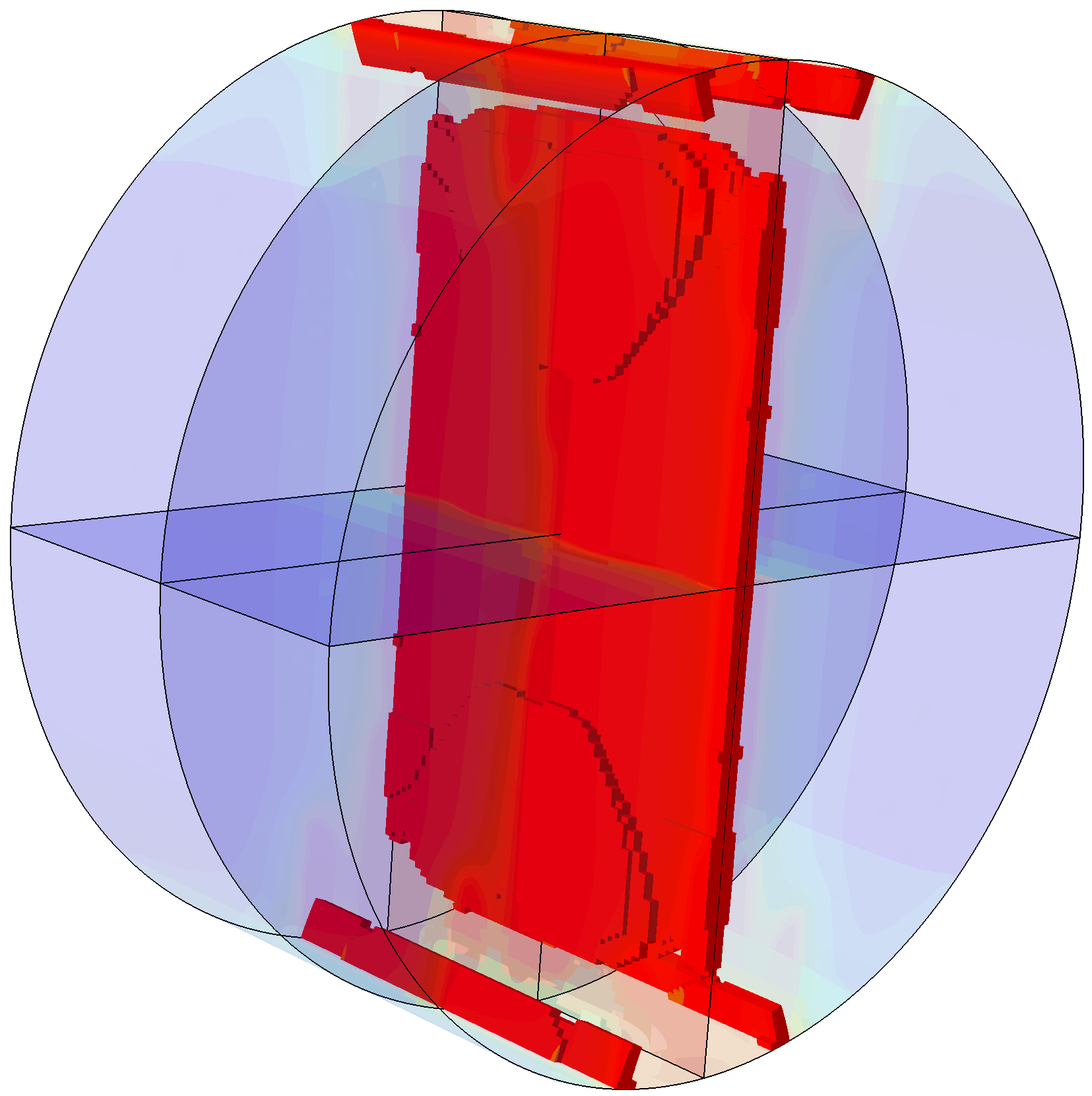}
        \caption{}
    \end{subfigure}\hspace{0.1\textwidth}
    \begin{subfigure}[H]{0.4\textwidth}
        \includegraphics[width=\textwidth]{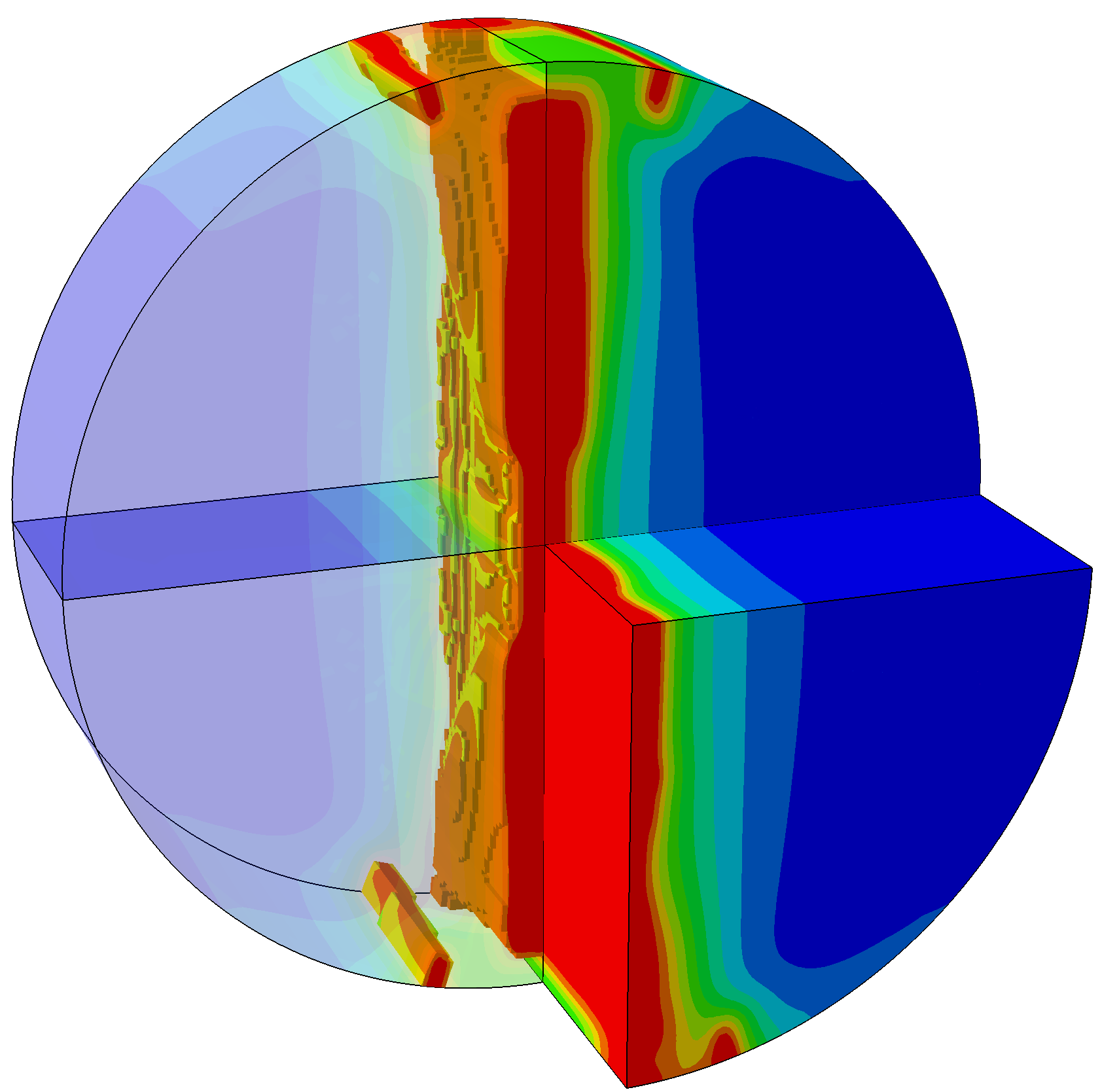}
        \caption{}
    \end{subfigure}
    \begin{subfigure}[H]{0.4\textwidth}
        \includegraphics[width=\textwidth]{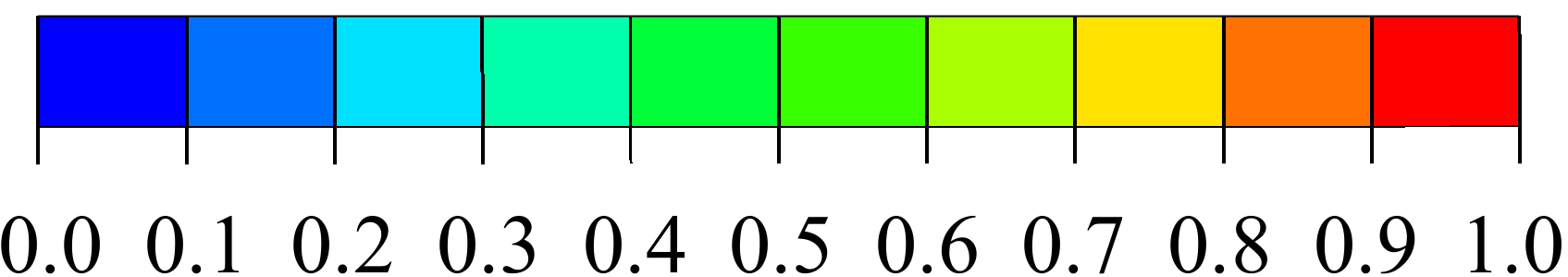}
    \end{subfigure}
    \begin{tikzpicture}[node distance=2cm]
    \node (cap) {{\Large $\phi$}};
    \end{tikzpicture}
    \caption{3D Brazilian test: contours of the phase field $\phi$ showcasing different stages of the fracture process. Sub-figures (a)-(c) show a transparent cross-section of the disk with $\phi > 0.9$ contours for the the following values of the remote displacement: (a) $u_y=-0.0668$ mm, (b) $u_y=-0.0670$ mm, and (c) $u_y=-0.0676$ mm. Sub-figure (d) shows the complete phase field $\phi$ contours for a jaw displacement of $u_y=-0.0676$ mm.}
    \label{Fig:BT-phi}
\end{figure}
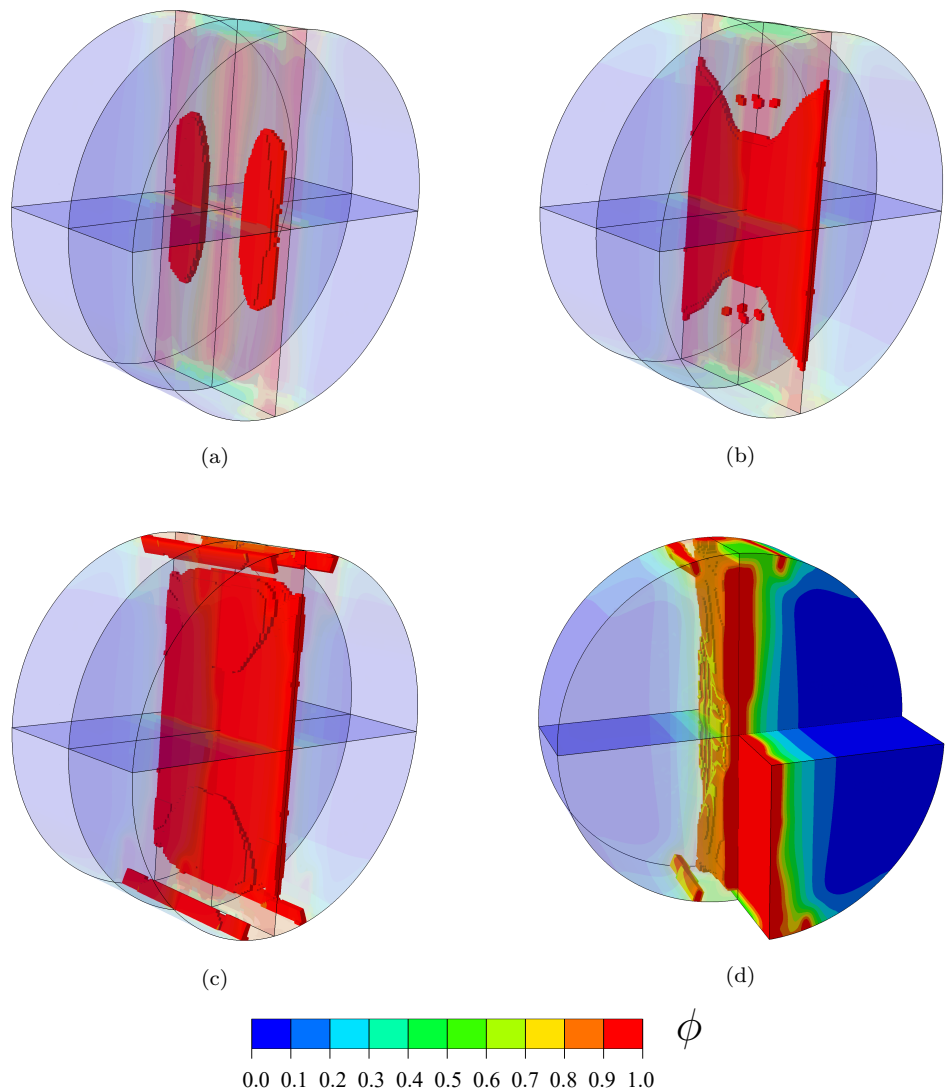

The results obtained are shown in Fig. \ref{Fig:BT-phi} in terms of the phase field contours for the different loading stages. The evolution of the phase field is also shown in Video 1, provided in the online version of this manuscript. Sub-figures \ref{Fig:BT-phi} (a)-(c) show in red colour the phase field contours where $\phi>0.9$. The crack appears to initiate at the centre of the disk and propagates towards the jaws very fast. Also, smaller cracks nucleate near the loading region. These calculations have been obtained using 345 load increments and using the monolithic implementation, no convergence issues were observed. 

\section{Conclusions}
\label{Sec:ConcludingRemarks}

We have presented a simple and robust implementation of the phase field fracture method in Abaqus. The framework developed does not require the coding of user-defined elements and therefore enables exploiting the majority of the in-built features of commercial finite element codes. This is achieved by taking advantage of the similarities between the heat transfer and the phase field evolution equations. The model can be developed entirely in Abaqus' graphical user interface and the implementation can be accomplished by combining a user material (UMAT) and a heat flux (HETVAL) subroutine. The code, which is provided open-source at www.empaneda.com/codes, can be used without changes for both 2D and 3D problems. The framework is general and can accommodate a wide variety of solution schemes and constitutive choices. Specifically, we incorporate both the spectral tension-compression \citep{Miehe2010a} and the volumetric-deviatoric \citep{Amor2009} strain energy decomposition. Moreover, we implement both monolithic and staggered solution schemes, providing a suitable trade-off between efficiency and robustness.\\

The potential of the framework is demonstrated by addressing four 2D and 3D paradigmatic boundary value problems. First, unstable fracture is examined using a notched square plate subjected to tension. Secondly, stable crack growth is investigated by subjecting the square plate to shear loading. Thirdly, the fracture of screws with and without internal cracks is investigated. Finally, the Brazilian test is simulated, including the modelling of the contact between the jaws and the disk. We observe that the monolithic standard Newton implementation provided is able to reach convergence in all cases. However, a single-pass staggered scheme appears to be more efficient in convergence-wise demanding problems. Computations are efficient but both schemes seem to perform worse than quasi-Newton methods \citep{Wu2020a,TAFM2020}. We also find that the use of interpolation schemes might not lead to efficiency improvements in phase field fracture. The framework can be very easily extended to other material models (e.g., plasticity) and damage mechanisms, such as fatigue. 

\section{Acknowledgements}
\label{Sec:Acknowledgeoffunding}

The authors would like to acknowledge financial support from the Ministry of Science, Innovation and Universities of Spain through grant PGC2018-099695-B-I00. E. Mart\'{\i}nez-Pa\~neda additionally acknowledges financial support from the EPSRC (grants EP/R010161/1 and EP/R017727/1) and from the Royal Commission for the 1851 Exhibition (RF496/2018).

\appendix

\section{Additional details of numerical implementation}
\label{App:FEM}

The framework can be easily extended to incorporate other constitutive choices. Specifically, as shown in the results section, a tension-compression split of the driving force for fracture should be considered to prevent damage from developing under compressive stresses. Alternative strain energy splits are described below, together with an anisotropic phase field formulation where the split is incorporated into the linear momentum equation. All these extensions are implemented in the user material (UMAT) subroutine. For simplicity, the code accompanying this manuscript (to be downloaded from www.empaneda.com/codes) does not include these additional features, but an extended version can be provided upon request.

\subsection{Strain energy density decomposition}

The two most widely used strain energy splits are considered: the \citet{Miehe2010a} tension-compression spectral decomposition and the \citet{Amor2009} volumetric-deviatoric split. In both cases, the strain energy density is decomposed as follows,
\begin{equation}
    \psi_0 = \left( 1 - \phi \right)^2 \psi_0^+ + \psi_0^- \, ,
\end{equation}

\noindent and only $\psi^+$ is considered in the evaluation of the history field $H$, Eq. (\ref{eq:History}). In regard to the specific constitutive definition of $\psi_0^+$, the volumetric-deviatoric split assumes that the compressive part of the volumetric strain energy does not contribute to the fracture process. Accordingly,
\begin{align}
     & \psi_0^+ = \frac{1}{2} K \langle \text{tr} \left( \bm{\varepsilon} \right) \rangle^2_+ + \mu \left( \bm{\varepsilon}' : \bm{\varepsilon}' \right) \\
     & \psi_0^- = \frac{1}{2} K \langle \text{tr} \left( \bm{\varepsilon} \right) \rangle^2_-
\end{align}

\noindent where $K$ is the bulk modulus, $\mu$ is the shear modulus, $\langle \rangle$ denote the Macaulay brackets, such that $\langle a \rangle_{\pm}=(a\pm |a|)/2$, and $\bm{\varepsilon}'$ is the deviatoric part of the strain tensor, such that $\bm{\varepsilon}'=\bm{\varepsilon}-tr(\bm{\varepsilon}) \bm{1} /3$. Here, $\bm{1}$ is the second-order unit tensor.\\

On the other hand, the spectral decomposition considers,
\begin{align}
     & \psi_0^+ = \frac{1}{2} \lambda \langle \text{tr} \left( \bm{\varepsilon}^+ \right) \rangle^2 + \mu \, \text{tr} \left[ \left( \bm{\varepsilon}^+ \right)^2 \right] \\
     & \psi_0^- = \frac{1}{2} \lambda \langle \text{tr} \left( \bm{\varepsilon}^- \right) \rangle^2 + \mu \, \text{tr} \left[ \left( \bm{\varepsilon}^- \right)^2 \right]
\end{align}

\noindent where $\lambda$ is the first Lamé constant and a spectral decomposition is applied to the strain tensor, such that:
\begin{equation}\label{eq:StrainSpectral}
    \bm{\varepsilon} = \sum_{I=1}^3 \langle \varepsilon_I \rangle \, \bm{n}_I \otimes \bm{n}_I 
\end{equation}

\noindent where $\varepsilon_I$ and $\bm{n}_I$ are the principal strains and principal strain directions (with $I=1,2,3$). The components $\bm{\varepsilon}^+$ and $\bm{\varepsilon}^-$ are obtained by considering in (\ref{eq:StrainSpectral}) the tensile and compressive principal strains, respectively.

\subsection{Anisotropic formulation}\label{Anisotropic formulation}

While the majority of the representative results presented are obtained using the hybrid approach proposed by \citet{Ambati2015}, we have also extended our implementation to incorporate the so-called \emph{anisotropic} approach \citep{Miehe2010a}. Thus, the decomposition into tension and compression components is also considered in the field equation for the displacement problem, such that the Cauchy stress (\ref{eq:Cauchy}) would instead read,
\begin{equation}
    \bm{\sigma} = \left( 1 - \phi \right)^2 \bm{\sigma}_0 = \left( 1 - \phi \right)^2 \frac{\partial \psi_0^+ \left( \bm{\varepsilon} \right) }{\partial \bm{\varepsilon}} + \frac{\partial \psi_0^- \left( \bm{\varepsilon} \right) }{\partial \bm{\varepsilon}}
\end{equation}

From an implementation perspective, this translates into a more elaborate computation of the material Jacobian, $\bm{C}=\partial \bm{\sigma}/ \partial \bm{\varepsilon}$. Thus, the material behaviour is characterised by the following 4th order elasticity tensor:
\begin{equation}
  \boldsymbol{C}=\lambda\left\{\left[(1-\phi)^{2}\right] {H}_{\varepsilon}(\text{tr}(\boldsymbol{\varepsilon}))+H_{\varepsilon}(-\operatorname{tr}(\boldsymbol{\varepsilon}))\right\} \boldsymbol{J} + 2 \mu \left\{\left[(1-\phi)^{2} \right] \boldsymbol{P}^{+}+\boldsymbol{P}^{-}\right\}  
\end{equation}

\noindent where ${H}_{\varepsilon}$ is the Heaviside function, such that $H_{\varepsilon}(x)=1$ for $x \geq 0$ or $H_{\varepsilon}(x)=0$ for $x<0$, and $\bm{J} \equiv J_{ijkl} = \delta_{ij} \delta_{kl}$, with $\delta_{ij}$ being the Kronecker delta. Also, the projection tensor $\bm{P}^+=\partial_{\bm{\varepsilon}} \left[ \bm{\varepsilon}_+ \left( \bm{\varepsilon} \right) \right]$ is computed as \citep{Miehe1998}
\begin{equation}\label{eq:Norma}
\begin{aligned}
P_{i j k l}^+ = & \sum_{a=1}^{3} \sum_{b=1}^{3} H_{\varepsilon} \left(\varepsilon_{a}\right) \delta_{a b} n_{a i} n_{a j} n_{b k} n_{b l} \\
& + \sum_{a=1}^{3} \sum_{b \neq a}^{3} \frac{1}{2} \frac{\left\langle\varepsilon_{a}\right\rangle_{+}-\left\langle\varepsilon_{b}\right\rangle_{+}}{\varepsilon_{a}-\varepsilon_{b}} n_{a i} n_{b j}\left(n_{a k} n_{b l}+n_{b k} n_{a l}\right) 
\end{aligned}
\end{equation}

\noindent where $n_{x i}$ is the $i^{th}$ component of the principal strain directions vector $n_x$. On the other hand: $\bm{P}^-=\bm{I}-\bm{P}^+$, with $\bm{I}$ being the fourth-order identity tensor. If $\varepsilon_{a}=\varepsilon_{b}$ then $P_{i j k l}^{+}$ (\ref{eq:Norma}) cannot be evaluated. Under such circumstances we replace the term $\left( \left\langle\varepsilon_{a}\right\rangle_{+}-\left\langle\varepsilon_{b}\right\rangle_{+} \right)/\left( \varepsilon_{a}-\varepsilon_{b} \right)$ with $H_{\varepsilon} \left(\varepsilon_{a}\right)$.



\bibliographystyle{elsarticle-harv}
\bibliography{library}

\end{document}